\documentclass[a4paper,11pt,final]{article}

\usepackage{authblk}

\usepackage{graphics}
\usepackage{graphicx}
\usepackage{epsfig}
\usepackage{amssymb}
\usepackage{amsthm}

\usepackage{amsmath}
\date{}

\begin{document}

\title{Stochastic foundations of
undulatory transport phenomena: Generalized Poisson-Kac processes - Part I Basic theory}

\author[1]{Massimiliano Giona$^*$}
\author[2]{Antonio Brasiello}
\author[3]{Silvestro Crescitelli}
\affil[1]{Dipartimento di Ingegneria Chimica DICMA
Facolt\`{a} di Ingegneria, La Sapienza Universit\`{a} di Roma
via Eudossiana 18, 00184, Roma, Italy  \authorcr
$^*$  Email: massimiliano.giona@uniroma1.it}

\affil[2]{Dipartimento di Ingegneria Industriale
Universit\`{a} degli Studi di Salerno
via Giovanni Paolo II 132, 84084 Fisciano (SA), Italy}

\affil[3]{Dipartimento di Ingegneria Chimica,
 dei Materiali e della Produzione Industriale
Universit\`{a} degli Studi di Napoli ``Federico II''
piazzale Tecchio 80, 80125 Napoli, Italy}
\maketitle

\begin{abstract}
This article introduces the notion of Generalized Poisson-Kac (GPK)
processes which generalize the  class of  ``telegrapher's noise dynamics''
 introduced
by Marc Kac \cite{kac} in 1974, using
Poissonian stochastic perturbations.  In GPK processes
the stochastic perturbation acts as a
switching amongst a set of stochastic velocity vectors
controlled by a Markov-chain dynamics.
GPK processes possess trajectory  regularity  (almost everywhere)
and  asymptotic
Kac limit, namely the convergence towards Brownian motion (and to
stochastic dynamics driven by Wiener perturbations),
which characterizes also the long-term/long-distance properties
of these processes. 
In this article we introduce the structural properties
of GPK processes, leaving all the physical implications
to part II and part III \cite{part2,part3}.
\end{abstract}

%
%

\section{Introduction}
\label{sec_1}

Statistical physics, nonrelativistic quantum mechanics and
field theory are profoundly intertwined with the concept
of Brownian motion and with the application of Langevin 
stochastic equations driven by Wiener processes \cite{gen1,gen2,gen3}. 
Starting
from the work of R. Feynman it is difficult to figure
out a compact formulation and application of quantum mechanics
without the reference to path-integrals over Brownian-Wiener
trajectories \cite{path1,path2}.

In statistical physics of gases, liquids, polymers
and colloids (soft-matter physics, for short), the use
of Brownian motion and Wiener processes can be regarded
as the legacy of a {\em large-number ansatz}, wherein the influence
of a manifold of small contributions, uniquely
characterized by their mean and variance justifies
the application of  stochastic perturbations
characterized by uncorrelated increments distributed in a normal
way, which is precisely the definition of a Wiener process 
\cite{chandra,ottinger,soft}.

The statistical characterization of microdynamic equations
written in the form of Wiener-Langevin stochastic dynamics
leads (using any representation of the stochastic integrals
in the meaning of Ito, Stratonovich, Klimontovich, etc.)
to parabolic  (forward Fokker-Planck) equations
for the probability density function, in which
a deterministic drift ${\bf v}({\bf x})$,
and a tensorial diffusivity ${\mathbb D}$ can be always
identified, possessing the
structure of a second-order advection-diffusion equation.

The analogy  between Fokker-Planck equations
and advection-diffusion problems permits to identify
the overall probability flux ${\bf J}_p({\bf x},t)$ associated
with the probability density $p({\bf x},t)$ as
\begin{equation}
{\bf J}_p({\bf x},t) = {\bf v}({\bf x}) \, p({\bf x},t) - {\mathbb D}
\, \nabla p({\bf x},t)= {\bf J}_{p,c}({\bf x},t) + {\bf J}_{p,d}({\bf x},t)
\label{eq_1_1}
\end{equation}
where the convective contribution ${\bf J}_{p,c}({\bf x},t)$
refers to the mean velocity drift ${\bf v}({\bf x})$
and the diffusive contribution ${\bf J}_{p,d}({\bf x},t)$
accounts for  the action of the stochastic perturbations, and is
proportional to the probability concentration gradient
$\nabla p({\bf x},t)$ multiplied by the tensorial
diffusivity ${\mathbb D}$, which corresponds to a generalized
Fickian relation.

In order words, there is a one-to-one correspondence between
Wiener-modelling of fluctuations at the microscopic level
and Fickian constitutive equations. This correspondence, starting
from the microscopic Wiener paradigm (i.e., from the  stochastic
Langevin equations driven by Wiener perturbations for
micro/mesoscopic particle motion) reflects itself also in the hydrodynamic
limit of these equations, i.e., in the formulation 
of the classical theory of irreversible process, such as
embodied in the monographs by de Groot and Mazur \cite{de_groot}
and Prigogine \cite{prigogine}, wherein the
 second principle of thermodynamics - developed for continua
upon a preliminary identification of the entropy production rates
associated with the different irreversible processes -
is naturally fulfilled
by constitutive equations of Fickian nature.

The Wiener paradigm, and its macroscopic  counterpart 
expressed by constitutive equations of 
Fickian type, imply several major physical
issues:
\begin{itemize}
\item A generic realization of a Wiener process admits
a trajectory that, with probability $1$, is an almost
nowhere rectifiable curve (in point of fact, it is a fractal
curve possessing Hausdorff dimension 3/2) \cite{falconer}. 
Consequently, with
probability
$1$, the velocity is unbounded, which contrasts the
constraint imposed by special relativity to
the propagation of physical particles and fields, limiting
the application of the theory to timescales much larger than
fluctuational timescale.
\item A parabolic advection-diffusion equation, in which
the diffusive flux admits a Fickian constitutive equation
possesses an infinite propagation velocity, again in striking
contraddiction with special relativity.
\item The parabolic nature of the Fokker-Planck equations 
associated with Wiener driven stochastic microdynamics
cannot be consistently embedded in the space-time description
of special relativity as the time coordinate enters as
a first-order partial derivative and spatial coordinate
(due to the Laplacian contribution accounting for effect of
the diffusive Fickian  flux), as second-order partial derivative.
\end{itemize}
These issues, are indeed formal manifestations  of one and the
same physical problem that is entirely associated
with the simplifying coarse-grained assumptions underlying
the statistics of Wiener processes, determining their fractal,
almost-everywhere non-differentiable, character.

It is important to stress that the lack of  relativistic
consistency of the classical
parabolic paradigm of continuum dynamics of irreversible processes
is crucial not because we are considering  high-energy transport processes,
for which the velocities  can approach that of
light {\em in vacuo}, but essentially because, 
 classical transport models and the  thermodynamics of irreversible phenomena
are regarded as a phenomenological, ``second-rank'', theories with respect
to the conservative theoretical formulations of field theory (both
classical and quantum), in the sense that the former are viewed
as corse-grained approximations of the latter, 
as in the Zwanzig projection method \cite{zwanzig}.

In 1948 and in subsequent works C. Cattaneo \cite{cattaneo1,cattaneo2}
and subsequently
 P. Vernotte \cite{vernotte}
proposed a hyperbolic diffusion model for heat conduction,
 based on a constitutive equation with memory,  capable of
solving the issue of infinite propagation velocity associated
with the classical parabolic heat equation. In the
Cattaneo model, the constitutive equation for
the diffusive flux (in the framework of heat transfer, this
flux corresponds to the generalization of the Fourier conductive
contribution),
 becomes
\begin{equation}
\tau_c \, \partial_t {\bf J}_d({\bf x},t) + {\bf J}_d({\bf x},t)
= - D \, \nabla p({\bf x},t)
\label{eq_1_2}
\end{equation}
where $\tau_c$ is a characteristic relaxation time that,
once inserted into the balance equation $\partial_t p({\bf x},t)=-\nabla
\cdot {\bf J}_d({\bf x},t)$, provides the hyperbolic diffusion
equation
\begin{equation}
\tau_c \, \partial_t^2 p({\bf x},t) + \partial_t p({\bf x},t)
= D \nabla^2 p({\bf x},t)
\label{eq_1_3}
\end{equation}
characterized by a finite propagation velocity.
Starting from this model, many different hyperbolic
transport equations have been proposed \cite{heat1,heat2,heat30,heat3,heat4,heat5},
 by modifying
slightly the formal structure of the Cattaneo equation (\ref{eq_1_3}),
and applied to a manifold of different physical phenomenologies \cite{catappl1,catappl2,catappl3,catappl4}.

The statistical (kinetic) derivation of the hyperbolic Cattaneo model
is essentially based upon  the Taylor series 
 expansion of the Chapman-Kolmogorov
equation for the probability density function, up
to the second order both in time and space variables \cite{statder1,statder2}.
As any Taylor series expansion and truncation of the evolution
equation for the probability density
function of a  Markovian process, this procedure does not
ensure that the solution of the truncated balance equation
would satisfy positivity requirements, i.e., that,
starting from an initial  non-negative probability density, the
solution of the truncated equation remains non-negative at any positive time
instant.
This issue follows straightforwardly as a consequence of the Pawula theorem
\cite{pawula}, and is further addressed below.
A recent critical analysis of the statistical derivation of the
Cattaneo equation is developed by Zhang et al. \cite{statder3}.
Both \cite{statder2,statder3} contain references on the first
statistical and kinetic attempts to derive the Cattaneo hyperbolic model.
The article by Metzler and Compte \cite{statder4} develops the derivation
of Cattaneo-type transport models from Continuous Time Random Walk.
The analysis is limited to one-dimensional spatial problems, but the
Cattaneo model is also generalized to the anomalous case involving
fractional time derivatives.
 
The Cattaneo paradigm, coupled to an extensive analysis of the
Grad's 13-moment expansion of the collisional Boltzmann equation \cite{grad},
are   the two main conceptual building blocks
that inspired and motivated a generalized theory of irreversible phenomena,
 initiated
by I. M\"uller and T. Ruggeri \cite{muller1,muller2,muller3}
 and subsequently elaborated
by many  researchers (D. Jou, D. Lebon, A. Casas-Vazquez, and many others)
\cite{jou0,jou1,jou2,jou3}
and referred to as {\em extended thermodynamics}.

The main conceptual difference between  extended thermodynamics
and the classical  theory of irreversible processes \cite{de_groot}
is the
generalization of the functional structure
of the   thermodynamic functions
of state in out-of-equilibrium conditions
that, in these theories,
can depend also on the  thermodynamic fluxes. In this framework the
Cattaneo model provides an instance of simple constitutive equation motivating
this generalization.
It is fully natural to ask for its microscopic interpretation
in terms of some underlying stochastic microdynamics.
A simple and beautiful answer to this question, in the case of one-dimensional
spatial problem, i.e., $x \in {\mathbb R}$, has been given by
M. Kac in 1974 \cite{kac}, that showed that the one-dimensional Cattaneo
equation can be regarded as the Fokker-Planck equation
associated with stochastic perturbations controlled
by Poisson processes of the form $(-1)^{\chi(t)}$,
where $\chi(t)$ is an ordinary Poisson process.
The ``telegrapher's noise'' considered by Kac  in \cite{kac}
is 
analogous to the model
of persistent random walk considered by S. Goldstein \cite{goldstein},
possessing an  exponentially decaying correlation function with time.

The Poisson-Kac perturbation $(-1)^{\chi(t)}$ corresponds to
a dichotomous noise, the  amplitude of which switches, with an exponential
distribution of switching times, between
the values $+1$ and $-1$. For this reason, the
body of literature originated from  Kac's article
has  used this model to study the effects of 
dichotomous, bounded noise on physical systems \cite{dicho1,dicho2,dicho3,dicho4,dicho5,dicho6,dicho7,dicho8,dicho9,dicho10,dicho11} or has 
applied this  stochastic system in order to address  the 
 influence  of colored noise on stochastic dynamics
 \cite{color1,color2,color3,color4}. For a review on dichotomous
noise see \cite{weiss,bena}.
Particularly relevant is the  analysis, within the Poisson-Kac paradigm, 
of  noise-induced transitions  \cite{horsthemke}, 
i.e.,  of phase-transitions controlled
by noise, occurring in a variety of physical
systems, from rheological models to chemical kinetics \cite{nitrans1,nitrans2,nitrans3,nitrans4}.

Using a continuation of the time variable towards the imaginary
axis,
Gaveau et al.   showed in one dimensional spatial models
that the statistical description of the simplest Poisson-Kac stochastic
dynamics 
corresponds to the  one-dimensional Dirac's equation
for the free electron \cite{gaveau}. This approach have  been subsequently
generalized by several authors \cite{dirac1,dirac2,dirac3}. 
Jona-Lasinio and coworkers developed several field-theoretical
calculations for fermions, including a Feynman-Kac theorem,
for stochastic quantized models using Poisson (and Poisson-Kac)
 processes \cite{jona1,jona2}.  

In point of fact, the correspondence between the Cattaneo
macroscopic model and Poisson-Kac processes holds exclusively
for one-dimensional space dynamics. More precisely in two- or
higher dimensions there is no stochastic process admitting
the Cattaneo hyperbolic
transport model as its forward Fokker-Planck equation.
The lack of microscopic stochastic explanation for the Cattaneo
model in higher dimensions ($n \geq 2$) follows directly
from the direct observation that the Green function associated
with eq. (\ref{eq_1_2}) for ${\bf x} \in {\mathbb R}^n$, $n \geq 2$,
attains negative values, as addressed in \cite{korner}.
This  observation raised doubts and criticisms
on higher-dimensional hyperbolic transport models,
and on the 
stochastic foundations of extended thermodynamics theories based upon
it  (see \cite{criticism0,criticism1} and  the analysis in  \cite{criticism_extended}). A similar result concerning negative solutions
for pulse propagation, i.e., for the Green function of
another hyperbolic model (the Guyer-Krumhansl model) has been
recently reported \cite{zhuk}.

Several works have been published on higher dimensional
extension of Poisson-Kac processes. Particularly
interesting is the analysis developed by Kolesnik and coworkers 
on two-dimensional generalizations, which leads, for the
overall probability density function, to extremely complicated
balance equations \cite{kolesnik1,kolesnik2}.
Concerning persistent random walks on a lattice, the 
works by Masoliver et al. \cite{masoliver1,masoliver2} and
by Boguna et al. \cite{boguna}
on two- and three-dimensional walk models showed that
their statistical description does not reduce to
a higher-dimensional Cattaneo equation but is slightly
more complex, see also \cite{colin}.

The aim of this article,
and of the forthcoming two, referred to as part II and part III
\cite{part2,part3}, respectively, is to  provide a stochastic
foundation for generic stochastic processes possessing finite
propagation velocity, and almost everywhere smooth trajectories -
generalizing the one-dimensional Poisson-Kac paradigm -
and to derive their properties and implications
in non-equilibrium thermodynamics, kinetic and transport theories 
\cite{rosenau}.
The class of processes we consider is referred to as Generalized
Poisson-Kac processes (GPK for short). A preliminary definition of GPK
processes can be found in \cite{giona_gpk}.

Due to the hyperbolic nature of the balance equations
for the vector-valued probability density function
describing statistically GPK processes, the associated
transport phenomena originating from microscopic GPK dynamics
can exhibit - depending on the values of the parameters
defining the specific GPK process - either relaxation
dynamics typical of parabolic dissipative models (e.g. the
usual diffusion equation) or more pronounced wave-like properties.
For this reason, the class of statistical and transport models
originating from GPK microdynamics will be referred to as
``undulatory transport processes'', just to highlight their
wave-like propagation.

 In this article,
and in the subsequent parts II and III, the theory of GPK
processes is thoroughly developed focusing on the most relevant
physical implications, which range from a stochastic interpretation
of the Boltzmann collisional equation to the stochastic backbone
for  extended thermodynamic approach of nonequilibrium processes.

The approach followed  in defining GPK processes is
 to embed the original Poisson-Kac model within
a Markovian transition structure  amongst
$N$ possible distinct states 
characterizing the stochastic perturbation
(we refer to this model a the finite $N$-state
Poisson process), which in turn modulates a finite number of possible
velocity perturbations.
This class of models is very simple and very flexible, and contains
Brownian motion (Wiener processes) as a limiting case (the Kac limit).
The almost everywhere smooth nature of the trajectories provides
the occurrence of a finite propagation velocity, and overcomes
the problems stated above for Wiener-driven processes.
Remarkably, the generalization of this class of models to
include nonlinear effects, and a continuum of stochastic
states leads directly to a stochastic model corresponding to
the classical collisional Boltzmann equation for a particle
gas. In this sense, GPK theory   strengthens and  completes the
original program by M. Kac in kinetic theory \cite{kacprog1,kacprog2,kacprog3}, to provide
a Markovian stochastic model for the Boltzmann kinetic
equation without using simplifying assumption, or toy-model 
representation for the collisions amongst the molecules.
Stemming from GPK theory, several byproducts follow, e.g.,
the Poisson-Kac mollification 
of Wiener processes that can be useful in analysis of  stochastic
partial differential equations and field theory.

While all the extensions and physical applications are addressed
in parts II and III, the present particle focuses almost exclusively
on the formal setting of GPK processes and on its basic
properties as regards the convergence towards Wiener-driven
Langevin equations (the Kac limit).
The article is organized as follows. All the basic 
definitions and properties of the classical one-dimensional Poisson-Kac
model, useful for the proper framing of GPK theory,
are reviewed in the Appendix at the end of the article.
Section \ref{sec_2} formalizes the guiding principles
underlying the development of Generalized  Poisson-Kac processes.
Section \ref{sec_3} addresses the multi-dichotomic extension
of the original Kac model in higher dimensions and its formal
drawbacks, essentially due to its ``heavy'' formalism and  due 
to the difficulty
in performing a continuum-extension of the model. These
difficulties
justify the introduction of Generalized Poisson-Kac
processes addressed in Section \ref{sec_4}.
In GPK processes, 
the driving stochastic mechanism is a
$N$-state finite Poisson process, modulating the occurrence 
of a given velocity vector out  of a family
of  $N$ stochastic velocity vectors.
The statistical description
of this class of processes involves $N$ partial probability
density  waves,
the evolution of which is controlled by a family of first-order
hyperbolic equation with recombination. The ``structural theory''
of GPK processes, i.e., the dynamic properties associated with
 the choice of the  transition rates,
the transition probability matrix, and the stochastic velocity vectors
is developed. This structural theory  bears  some resemblance with
similar calculations lying in the background of Lattice
Boltzmann models \cite{lattice_boltzmann}, and
with the theory of the discrete Boltzmann equation
\cite{dbe1,dbe2,dbe3}. Section \ref{sec_5} focuses on 
the Kac limit of
GPK processes, i.e., on the convergence of  this class
of stochastic models towards Brownian motion and Wiener-driven Langevin
equations, whenever the intensity
of the stochastic velocity vectors $b^{(c)}$ and the  characteristic
transition rate
$\lambda^{(c)}$
diverge to infinity, keeping fixed the ratio $(b^{(c)})^2/2 \lambda^{(c)}$.
Homogenization theory of GPK processes is addressed in Section \ref{sec_6},
showing that, under certain conditions, the Kac limit
can be viewed as a long-term emerging property of this
class of models.

\section{Basic principles}
\label{sec_2}

The structure of  one-dimensional Poisson-Kac processes, reviewed in the
Appendix, contains the germs for its generalization
in higher dimensions. Three basic principles can be enucleated
out of  it, that can guide  the
 development of a simple and consistent stochastic theory of undulatory
transport phenomena. These principles are:
\begin{itemize}
\item The principle of stochastic reality;
\item The principle of the primitive variables;
\item The principle of the  asymptotic Kac convergence.
\end{itemize}
Below, their meaning and importance is outlined and discussed.

\subsection{The principle of stochastic reality}
\label{sec_2_1}

The principle of stochastic reality states that all the
 transport models should be derived
from an underlying microscopic equation of motion.

The evolution equation for the probability density function
follows from the structure of the stochastic microdynamics.
Moreover, from the analysis of the
moments associated with the probability density function,
it is possible to derive the transport equations, analogously
to what usually applied in kinetic theory (using the
 5-moment recipe of classical irreversible thermodynamics,
or the Grad's 13-moment expansion).

In undulatory transport theory, the microdynamics can be
written as a generalization of the Poisson-Kac
model reviewed in the Appendix.
In  the statistical theory of Poisson-Kac processes,
 the equivalent of the Fokker-Planck equation (associated with
 classical
Wiener-driven Langevin equations)
is given by the system of two first-order equations (\ref{eq2_8}),
involving the   partial probability densities $\{p^+(x,t), p^-(x,t)
\}$.  This observation  leads further
 to the second principle we adopt, namely that
of primitive variables, introduced and discussed 
in the next paragraph.

The principle of stochastic reality admits two fundamental
implications. To begin with, it automatically
ensures that the transport equations for scalar concentration
fields (such as molar concentrations of chemical species,
or the absolute temperature),
strictly non-negative quantities by definition, would  automatically fulfill
this fundamental consistency requirement (at least, choosing
in a consistent wave-like way the boundary conditions, whenever
transport problems in bounded domains are considered \cite{brasiello}).
Moreover, it implies that, given a statistical description
for the evolution of a scalar concentration/probability field,
it is always possible to obtain a trajectory-based (Lagrangian) description
for the granular entities (atoms, molecules, particles, aggregates, etc.),
the statistical description of which constitutes the
essence of the transport problem under examination.

\subsection{The principle of primitive variables}
\label{sec_2_2}

The formulation of the statistical description of the
microscopic dynamics leads automatically to the
identification of the primitive fundamental
statistical variables describing the
process.
In stochastic dynamics associated with Wiener-driven Langevin equations,
$d {\bf x}(t) = {\bf v}({\bf x}(t)) \, dt + \sqrt{2 D} d {\bf w}(t)$,
where $D$ is a constant diffusivity, ${\bf x}=(x_1,\dots,x_n)$,
and ${\bf w}(t)=(w_1(t),\dots,w_n(t))$ is a $n$-dimensional vector-valued
Wiener process,
the primitive variables are obviously the
probability density function $p({\bf x},t)$ and its diffusive
flux ${\bf J}_d({\bf x},t)$. The
balance equation is of the form $\partial_t p({\bf x},t)= -\nabla \cdot \left [ {\bf v}({\bf x}) \, p({\bf x},t) \right ] - \nabla \cdot {\bf J}_d({\bf x},t)$,
and the influence of a stochastic perturbation, possessing normally-distributed 
independent increments, automatically defines a constitutive equation,
relating the diffusive flux to the concentration gradient
of $p({\bf x},t)$, 
\begin{equation}
{\bf J}_d({\bf x},t) = - D \, \nabla p({\bf x},t)
\label{eq3_1}
\end{equation}
of Fickian nature. This is because Wiener fluctuations are completely
``renormalized'' out of the Fokker-Planck equation,
and this complete renormalization implies a Fickian-type,
memoryless,
constitutive equation.

Re-analyzing the one-dimensional model discussed
in the Appendix, a one-dimensional Poisson-Kac process can  still be
described in terms of an overall probability density $p(x,t)$, and
of its diffusive flux $J_d(x,t)$.
However, just because of the finite-propagation of the stochastic
perturbation, the flux-concentration description is a derived
byproduct of the fundamental description of the process, that
is naturally expressed by means of the  two partial probability 
waves $p^+(x,t)$,
$p^-(x,t)$.

The primitive variables of the one-dimensional Poisson-Kac process
(\ref{eq2_3}) are just $p^\pm(x,t)$. In point of fact, the
statistical description of the process  in terms
of the partial probability waves, eq. (\ref{eq2_8}), is definitely
 simpler 
than   the corresponding
concentration/flux $(p,J_d)$-description  based on eq. (\ref{eq2_10}).

Expressed in terms of the primitive variables, boundary conditions
becomes much simpler to define and enforce,
even in the one-dimensional spatial case \cite{brasiello}.

Therefore, in the development of a undulatory theory of transport
processes, the following identification principle should be
fruitfully used: {\em The statistical structure of a 
stochastic perturbation naturally induces a system of 
primitive statistical variables. The definition of transport
equations and boundary conditions should involve the properties
(e.g. the moments) associated with these primitive variables,
out of which overall concentrations, and overall fluxes can be
determined as derived quantities}.

As we will see in the next Sections, and further in part II and III,
 in higher-dimensional
transport problems the classical paradigm based on  an overall
concentration and its associated ``diffusive''
 flux breaks down completely and becomes 
useless both for theoretical
development and for practical purposes. Conversely,
 a suitable system of partial probability  waves
can always be defined, being fairly simple to handle both
in theoretical development and in numerical calculations.

It is rather clear that this radical shift in defining primitive statistical
variables changes completely the way transport equations are formulated
and, as a consequence, modify the formal structure of an extended irreversible
thermodynamic theory that can be built upon them.
A typical example supporting this claim occurs in the
formulation of entropy functions and in the formalization
of the second principle of thermodynamics see \cite{part2}.

To conclude, it may sound ``artificial'' that
 the overall concentration (overall probability
density) and its diffusive flux - which are the  physical quantities
easily amenable to a direct experimental measurement - are
regarded as derived quantities, while the primitive
observables of the theory are a system of partial probability
densities which are less intuitive to figure out. 
This situation is however rather common in physics.
In non-relativistic quantum  theory, the primitive variable
is the wave-function $\psi({\bf x},t)$, solution of the
Schr\"odinger equation, while the quantity of direct physical interest is
its square modulus $|\psi({\bf x},t)|^2$, that, using the Born
ansatz, defines a spatial probability density function for the
quantum system. The situation is even more clear, and conceptually
analogous to the use of partial waves, when the relativistic
quantum theory is consider, and the primitive quantities
of Dirac's theory are spinors, i.e. 4-dimensional field variables
that covariantly transform under a Lorentz boost. Indeed,
the connection between the principle of the primitive variables
and the spinorial formulation (in $L^1$, and not in $L^2$
as in the Dirac quantum theory) is strict \cite{giona_spinorial}.

\subsection{The principle of the  asymptotic Kac convergence}
\label{sec_2_3}

The principle of the  asymptotic Kac convergence is a closure
condition with respect to Brownian motion (Wiener process).
Given an undulatory stochastic dynamics
in the presence of a deterministic biasing field ${\bf v}({\bf x})$,
and let $b^{(c)}$ and $\lambda^{(c)}$, the characteristic
velocity associated with the propagation of stochastic fluctuations
and the characteristic transition rate, respectively (the expressions
for $b^{(c)}$, and $\lambda^{(c)}$ in the framework
of Generalized Poisson-Kac processes are developed in Section \ref{sec_5}).
In the case of eq. (\ref{eq2_9}), $b^{(c)}=b$, $\lambda^{(c)}=\lambda$.

If we let $b^{(c)}$ and $\lambda^{(c)}$ diverge,
keeping constant the ratio
\begin{equation}
\lim_{b^{(c)}, \lambda^{(c)} \rightarrow \infty}
\frac{(b^{(c)})^2}{2 \, \lambda^{(c)}} = D_{\rm eff}
\label{eq3_2}
\end{equation}
and equal to a diffusivity $D_{\rm eff}$ (Kac limit), then
the statistical description of the system would involve  solely
the overall probability density function $p({\bf x},t)$ that,
in the Kac limit,  should be  a solution of  the
parabolic advection-diffusion equation
\begin{equation}
\partial_t p({\bf x},t) = - \nabla \cdot \left [ {\bf v}({\bf x}) \, p({\bf x},t)
\right ] + D_{\rm eff} \, \nabla^2 p({\bf x},t)
\label{eq3_3}
\end{equation}

The Kac-limit principle can be interpreted in a three-fold way:
\begin{itemize}
\item It ensures that the classical parabolic models of transport
are limit cases of the corresponding undulatory counterparts
whenever the  characteristic velocity $b^{(c)}$ and transition
rate $\lambda^{(c)}$ of the
stochastic perturbations are no longer finite, still keeping fixed
the value of a functional relation  amongst them eq. (\ref{eq3_2});
\item It provides analytical criteria for deriving
basic constraints on the parameters entering the generalization
of Poisson-Kac  processes (see Section \ref{sec_5});
\item Mutuating the analogy with quantum mechanics, it represents
a form of ``semiclassical limit'' of all the undulatory
models of stochastic dynamics, with respect to their
Wiener-driven counterparts.
\end{itemize}

The rigorous assessment of the Kac limit for a generic
model of undulatory transport is not a simple mathematical
task, as analyzed by Kolesnik in some cases \cite{kolesnik3}.
However, a physically-oriented approach to the
Kac limit is developed in Section \ref{sec_6}, grounded
on an equipartition principles amongst the partial probability 
waves in their recombination dynamics.

\section{Multi-dichotomic extensions of the Poisson-Kac model}
\label{sec_3}

In this Section we analyze some extensions of the
Poisson-Kac model reviewed in the Appendix to
one- and higher-dimensional problems, based on a ``multi-dichotomic''
description of the stochastic perturbation.

\subsection{One-dimensional Poisson-Wiener mixed model}
\label{sec_3_1}

If one releases the assumption of finite
propagation velocity of the stochastic perturbation,
a ``mixed'' stochastic model can be considered for
systems subjected to both Wiener and Poisson perturbations.
Below, we analyze the one-dimensional case.
Consider the stochastic differential equation
\begin{equation}
d x(t) = v(x(t)) \, dt + b \, (-1)^{\chi(t)} \, dt + \sqrt{2 D_w} \, d w(t)
\label{eq4_1}
\end{equation}
where $d w(t)$ are the increments in the
interval $(t,t+dt)$ of a one-dimensional Wiener 
process, and $D_w>0$. Assume that $W(t)$ and $\chi(t)$
are independent of each other. The statistical
description of this model is still based on the
two partial probability densities $p^{\pm}(x,t)$ that,
in the present case, fulfill the system of parabolic
equations
\begin{eqnarray}
\partial_t p^+(x,t) & = & -\partial_x \left [ v_+(x) \, p^+(x,t) \right ]
+ D_w \, \partial_x^2 p^+(x,t)
-\lambda \, p^+(x,t) + \lambda \, p^-(x,t)
\nonumber \\
\partial_t p^-(x,t) & = & -\partial_x \left [ v_-(x) \, p^-(x,t) \right ]
+ D_w \, \partial_x^2 p^-(x,t) 
+\lambda \, p^+(x,t) - \lambda \, p^-(x,t) \nonumber \\
\label{eq4_2}
\end{eqnarray}
where $v_+(x)=v(x)+b$, and $v_-(x)=v(x)-b$.
In terms of $p=p^++p^-$ and $J_d= b \, (p^+-p^-)$, these
equations can be rewritten as
\begin{eqnarray}
\partial_t p  =  - \partial_x (v \, p) - \partial_x J_d +
D_w \, \partial_x^2 p \nonumber \\
\frac{1}{2 \, \lambda} \, \partial_t J_d + J_d  =  
- \frac{1}{2 \, \lambda} \left [ \partial (v \, J_d)
- D_w \, \partial_x^2 J_d \right ] - D_{\rm eff} \, \partial_x p
\label{eq4_3}
\end{eqnarray}
where $D_{\rm eff}= b^2/2 \lambda$. Eqs. (\ref{eq4_3}) represents the
one-dimensional, stochastically consistent, archetype of the
class of Guyer-Krumhansl models \cite{guyer1,guyer2}.

In the Kac limit, this model reduces to an advection-diffusion
equation with an effective diffusivity $D_{\rm tot}=D_{\rm eff}+D_w$,
which is the sum of the diffusivity $D_w$ associated with
the Wiener perturbation, and  $D_{\rm eff}$ deriving
from the Poissonian contribution.

\subsection{Higher-dimensional multi-dichotomic extensions}
\label{sec_3_2}

Using a system of dichotomic perturbations, it is rather
straightforward to develop higher-dimensional extensions
of the one-dimensional Poisson-Kac model reviewed in the Appendix.

To begin with, consider a two-dimensional case, $n=2$, letting
$\chi_1(t)$, $\chi_2(t)$ be two Poisson processes, independent
of each other, characterized by the
same transition rate $\lambda$. Consider the
stochastic dynamics
\begin{eqnarray}
d x_1(t) & = & v_1({\bf x}(t)) \, dt + b \, (-1)^{\chi_1(t)} \, dt
\nonumber \\
d x_2(t) & = & v_2({\bf x}(t)) \, d t + b \, (-1)^{\chi_2(t)} \, dt
\label{eq4_4}
\end{eqnarray}
and set ${\bf x}=(x_1,x_2)$, ${\bf v}({\bf x})=(v_1({\bf x}),v_2({\bf x}))$.
In the present case, the statistical description of the dynamics
involves a system of four partial probability density functions
$p^{(\alpha,\beta)}({\bf x},t)$, $\alpha,\beta=\pm$
\begin{equation}
\hspace{-1.5cm} p^{(\pm,\pm)}({\bf x},t) \, d {\bf x} =
\mbox{Prob} \left [ {\bf X}(t) \in ({\bf x}, {\bf x}+d {\bf x})\; ,
\;\; (-1)^{\chi_1(t)}= \pm 1, \; (-1)^{\chi_2(t)}= \pm 1
\right ]
\label{eq4_5}
\end{equation}
representing the primitive statistical variables of the model.
Henceforth, $d  {\bf x}$ will indicate the $n$-dimensional
measure element ($n=2$ in the present case), and ${\bf X}(t) \in ({\bf x},{\bf x}+d {\bf x})$
is the compact notation for $X_h(t) \in (x_h, x_h+d x_h)$, $h=1,\dots,n$.
The partial probability waves satisfy the balance equations
\begin{eqnarray}
\partial_t p^{(+,+)} & = & - \nabla \cdot \left [ {\bf v} + b \,
{\bf e}^{(+,+)} \, p^{(+,+)} \right ] - 2 \, \lambda \, p^{(+,+)}
+ \lambda ( p^{(+,-)}+p^{(-,+)}) \nonumber \\
\partial_t p^{(+,-)} & = & - \nabla \cdot \left [ {\bf v} + b \,
{\bf e}^{(+,-)} \, p^{(+,-)} \right ] - 2 \, \lambda \, p^{(+,-)}
+ \lambda ( p^{(+,+)}+p^{(-,-)})
\nonumber \\
\partial_t p^{(-,+)} & = & - \nabla \cdot \left [ {\bf v} + b \,
{\bf e}^{(-,+)} \, p^{(-,+)} \right ] - 2 \, \lambda \, p^{(-,+)}
+ \lambda ( p^{(+,+)}+p^{(-,-)}) \nonumber \\
\partial_t p^{(-,-)} & = & - \nabla \cdot \left [ {\bf v} + b \,
{\bf e}^{(-,-)} \, p^{(-,-)} \right ] - 2 \, \lambda \, p^{(-,-)}
+ \lambda ( p^{(+,-)}+p^{(-,+)} )
\nonumber\\
\label{eq4_6}
\end{eqnarray}
where
\begin{equation}
\hspace{-1.5cm} {\bf e}^{(+,+)}= \left (
\begin{array}{c}
1 \\
1
\end{array}
\right ) \; , \;\;
{\bf e}^{(+,-)}= \left (
\begin{array}{c}
1 \\
-1
\end{array}
\right ) \; , \;\;
{\bf e}^{(-,+)}= \left (
\begin{array}{c}
-1 \\
1
\end{array}
\right ) \; , \;\;
{\bf e}^{(-,-)}= \left (
\begin{array}{c}
-1 \\
-1
\end{array}
\right )
\label{eq4_7}
\end{equation}
This model corresponds, with minor modifications, to the problem
considered by Plykhin \cite{plykhin}.
Even in this  simple case, the importance
of the primitive statistical description  can be fully appreciated.
A simple, dimensional argument is enlightening. The
full statistical description of this model involves
four partial probability density functions $p^{(\pm,\pm)}({\bf x},t)$.
The overall probability density function $p({\bf x},t)$,
and the diffusive flux associated with the Poissonian stochastic
perturbation ${\bf J}_d({\bf x},t)$ are given by
\begin{equation}
p({\bf x},t)= \sum_{\alpha,\beta= \pm} p^{(\alpha,\beta)}({\bf x},t)
\; , \;\;\; {\bf J}_d({\bf x},t) =
 b \, \sum_{\alpha,\beta= \pm} {\bf e}^{(\alpha,\beta)} \,
 p^{(\alpha,\beta)}({\bf x},t)
\label{eq4_8}
\end{equation}
It represents a system of three functions: $p({\bf x},t)$ and the
two entries of ${\bf J}_d({\bf x},t)$, which cannot fully describe the
statistical properties of the system, at least in a simple way as
the first-order evolution equations for the four  primitive
statistical functions $p^{(\alpha,\beta)}({\bf x},t)$.

It is straightforward to develop a multi-dichotomic
extension in arbitrary spatial dimension $n$, and in the
presence of an arbitrary number $N$ of independent
Poissonian perturbations.
Consider a $n$-dimensional space ${\mathbb R}^n$, and ${\bf x}=(x_1,\dots,
x_n) \in {\mathbb R}^n$. Let $\chi_1(t),\dots,\chi_N(t)$ be a system
of $N$ Poisson processes, independent of each other,
and characterized, for simplicity, by the same transition
rate $\lambda$. Let $\{ {\bf b}_\alpha \}_{\alpha=1}^N$
be a system of $N$ constant velocity vectors in ${\mathbb R}^n$,
and consider the stochastic differential equation 
\begin{equation}
d {\bf x}(t) = {\bf v}({\bf x}(t)) \, dt + \sum_{\alpha=1}^N {\bf b}_\alpha
\, (-1)^{\chi_\alpha(t)} \, dt 
\label{eq4_9}
\end{equation}
Due to the dichotomic nature of each $(-1)^{\chi_\alpha(t)}$-perturbation, 
a system of $2^N$ partial probability density
functions is required for the full description of the process.
In order to formalize the statistical description of the
stochastic
dynamics (\ref{eq4_9}), let
\[
 \{\boldsymbol{\varepsilon}_k \}_{k=1}^{2^N}
= \{ \underbrace{(\pm 1,\dots, \pm1)}_{N} \}_{k=1}^{2^N}
\]
be the $2^N$ different strings, the entries of which attain
values $\pm 1$, corresponding to all the possible
states of the $N$-vector of Poissonian stochastic
perturbations
\[
\left ( (-1)^{\chi_1(t)}, \dots, (-1)^{\chi_N(t)} \right )
\]
and let $\sigma_\alpha$, $\alpha=1,\dots,N$  be the $N$ string transformations
$\sigma_\alpha({\boldsymbol \varepsilon}) = {\boldsymbol \varepsilon}^\prime$
defined as follows: if ${\boldsymbol \varepsilon}=(\varepsilon_1,\dots,
\varepsilon_N)$, $\varepsilon_h=\pm 1$, $h=1,\dots,N$, then
\begin{equation}
{\boldsymbol \varepsilon}^\prime= 
\sigma_\alpha({\boldsymbol \varepsilon})=
(\varepsilon_1, \varepsilon_{\alpha-1},
-\varepsilon_\alpha, \varepsilon_{\alpha+1},\dots,\varepsilon_N)
\label{eq4_10}
\end{equation}
The transformation $\sigma_\alpha$ corresponds to the transition
of $(-1)^{\chi_\alpha(t)}$ from $\pm 1$ to $\mp 1$.
Let $\varepsilon_{k,\alpha}$ be the $\alpha$-th entry of ${\boldsymbol
\varepsilon}_k$.

The statistical description of the stochastic dynamics
(\ref{eq4_9}) involves $2^N$ partial probability
density functions $p_{{\boldsymbol \varepsilon}_k}({\bf x},t)$,
$k=1,\dots 2^N$, fulfilling the system of $2^N$ first-order partial differential
equations
\begin{eqnarray}
\partial_t p_{{\boldsymbol \varepsilon}_k}({\bf x},t) & = &
 - \nabla \cdot \left [ \left ({\bf v}({\bf x}) + \sum_{\alpha=1}^N
{\bf b}_\alpha \, \varepsilon_{k,\alpha} \right )
\, p_{{\boldsymbol \varepsilon}_k}({\bf x},t)  \right ] \nonumber \\
& + & \lambda \, \sum_{\beta=1}^N \left [ p_{\sigma_\beta({\boldsymbol \varepsilon}_k)}({\bf x},t) - p_{{\boldsymbol \varepsilon}_k}({\bf x},t) \right ]
\label{eq4_11}
\end{eqnarray}
Albeit formally correct and of general validity, this higher
dimensional extension of the Poisson-Kac stochastic dynamics
is too complex for being physically appealing. Conceptually,
it is similar to the model considered by Kac in order to
provide a simplified description
of the Boltzmann  equation for a particle gas \cite{kacprog1}.
However, the drawbacks of this formalization are evident, and
can be summarized as follows:
\begin{itemize}
\item The multi-dichotomous nature of the stochastic
perturbation, implies a complex formalism for the
associated partial probability density functions, the dynamics
of which is given by eq. (\ref{eq4_11}). It hinders further development
of the theory.
\item The number of partial probability density functions necessary
to achieve a complete statistical description of the model grows exponentially
with respect to the number of independent Poissonian perturbations $N$
considered.
\item A further major theoretical drawback is that this model
is not suitable for developing a continuum-limit
extension of the stochastic perturbation, i.e., to consider,
instead of a finite number of $N$ of Poissonian contributions,
a continuum of stochastic forcing terms, see further part III.
\end{itemize}
All these problem can be overcome by using an alternative approach,
which is discussed in the next Section.

\section{Generalized Poisson-Kac processes}
\label{sec_4}

The straight multidimensional extension of dichotomous Poissonian
perturbations discussed in Section \ref{sec_3} is not the
most convenient and physically tractable generalization
of Poisson-Kac stochastic dynamics, essentially because
 the number of primitive statistical descriptors
of this process grows exponentially as ${\mathcal O}(e^{N \, \log 2})$
as a function of the number $N$ of distinct and independent
Poissonian perturbations considered. Moreover, it is
fairly cumbersome to develop a continuum-extension of the stochastic
forcing.
An alternative approach is to consider a Markov-chain formulation
of the stochastic forcing, leading to
a class of processes that we refer to as {\em Generalized Poisson-Kac
processes} \cite{giona_gpk}.

The stochastic building block of this class of processes is represented
by the {\em $N$-state finite Poisson process} $\chi_N(t)$,
which is a stationary, memoryless, ordinary stochastic process
attaining $N$ distinct values $\{1,\dots,N\}$.

A $N$-state finite Poisson process is a Markov chain amongst $N$ states,
defined by a constant vector ${\boldsymbol \Lambda}=(\lambda_1,\dots,\lambda_N)$
of transition rates, $\lambda_\alpha>0$, $\alpha=1,\dots,N$, and by a constant 
matrix ${\bf A} = (A_{\alpha,\beta})_{\alpha,\beta=1}^N
 \in {\mathcal M}_{N \times N}$ of transition
probabilities. In the interval $(t,t+\Delta t)$ the transition from
$\chi_N(t)=\alpha$, to $\chi_N(t+\Delta t)=\beta$,
$\alpha,\beta=1,\dots,N$, is expressed by the
transition probabilities
\begin{eqnarray}
T_{\alpha \rightarrow \beta}(\Delta t) & =  &
\lambda_\alpha \, A_{\beta,\alpha}
\, \Delta t + o(\Delta t) \label{eq5_1} \\
T_{\alpha \rightarrow \alpha}(\Delta t) & = &
\left [ 1- \lambda_\alpha \, \sum_{\beta=1}^N A_{\beta,\alpha} \, \Delta t
\right ] + o(\Delta t) \nonumber
\end{eqnarray}
The ordinary nature of the process implies that the probability of having 
more
than a single transition in the interval $(t,t+\Delta t)$
is order of $o(\Delta t)$.
As regards the transition  probability matrix ${\bf A}$, it follows from
its probabilistic meaning that
\begin{equation}
A_{\alpha,\beta} \geq 0 \; , \;\;\;
\sum_{\alpha=1}^N A_{\alpha,\beta}=1 \;\; \; \; \beta=1,\dots,N
\label{eq5_2}
\end{equation}
i.e., that $A_{\alpha,\beta}$ is a left-stochastic matrix \cite{seneta}. 

Let $\widehat{P}_{\alpha,\beta}(t)$ be the conditional probability
of the occurrence $\chi_N(t)=\alpha$ given the initial
condition $\chi_N(0)=\beta$. From eqs. (\ref{eq5_1}), it follows
that $\widehat{P}_{\alpha,\beta}(t)$, $\alpha,\beta=1,\dots,N$ satisfy
the system of differential equations for the associated Markov chain
\begin{equation}
\frac{d \widehat{P}_{\alpha,\beta}(t)}{d t} =
- \lambda_\alpha \, \widehat{P}_{\alpha,\beta}(t)
+ \sum_{\gamma=1}^N \lambda_\gamma \, A_{\alpha,\gamma}  \widehat{P}_{\gamma,\beta}(t)
\label{eq5_3}
\end{equation}
Indicating with $P_\alpha(t)= \sum_{\beta=1}^N \widehat{P}_{\alpha,\beta}(t)$
the partial probabilities obtained by summing $\widehat{P}_{\alpha,\beta}(t)$
over all the possible initial states, it follows from
eq. (\ref{eq5_3}) that $P_{\alpha}(t)$, $\alpha=1,\dots,N$,
 satisfy eq. (\ref{eq5_3})
simply by replacing $\widehat{P}_{\alpha,\beta}(t)$ with $P_\alpha(t)$.

Let ${\mathcal B}_N=\{{\bf b}_{\alpha} \}_{\alpha=1}^N$ be a system of
$N$ constant (velocity) vectors in ${\mathbb R}^n$.
A {\em Generalized Poisson-Kac process} in ${\mathbb R}^n$,
in the presence of a deterministic velocity field ${\bf v}({\bf x})$,
is defined by the stochastic differential equation
\begin{equation}
d {\bf x}(t) =  {\bf v}({\bf x}(t)) \, dt + {\bf b}_{\chi_N(t)} \, d t
\label{eq5_4}
\end{equation}
where $\chi_N(t)$ is a $N$-state  finite Poisson process,
equipped with the initial conditions ${\bf x}(t_0=0)={\bf x}_0$.
Therefore a GPK is defined by the sextuple 
\[
(n,N,{\boldsymbol \Lambda},
{\bf A}, {\mathcal B}_N,{\bf v}({\bf x}))
\]
 where $n$ is
the dimensionality of the system variables ${\bf x} \in {\mathbb R}^n$
, $N$ is
the dimensionality of the stochastic perturbation, corresponding to the
number of states of the Markov chain defining the finite Poisson process
$\chi_N(t)$, ${\boldsymbol \Lambda}=(\lambda_\alpha)_{\alpha=1}^N$,
$\lambda_\alpha>0$ is the vector of the transition rates,
${\bf A} \in {\mathcal M}_{N\times N}$, a left stochastic matrix accounting
for the allowed transitions amongst the $N$ states, ${\mathcal B}_N=\{{\bf b}_{\alpha} \}_{\alpha=1}^N$, is a system of $N$ constant vectors corresponding to the possible stochastic
perturbations allowed in the system, and ${\bf v}({\bf x})$ is
a smooth vector field in ${\mathbb R}^n$, describing the deterministic
component of the dynamics.

Let us discuss in greater detail the properties of the parameters
defining a GPK.
The transition matrix ${\bf A}$ accounts for the states that
can be reached from any other states, and their transition probabilities.
It is rather intuitive to assume a form of {\em ergodicity},
that for the $N$-state finite Poisson process implies that
the left-stochastic matrix ${\bf A}$ is irreducible \cite{seneta}.
 This means
that if $P_\alpha(t=0)=\delta_{\alpha,\beta}$, for any $\beta=1,\dots,N$,
 then there exists
a time $t_\beta^*$, such that for $t>t_\beta^*$, $P_{\gamma}(t)>0$,
i.e.,  starting from any initial state $\beta=1,\dots,N$, 
any other state is reached in finite time.
Irreducibility implies the existence of a unique real Frobenius
eigenvalue $\mu_A=1$ of ${\bf A}$,  corresponding to the eigenvalue of ${\bf A}$ possessing maximum modulus.

 Using a form of detailed balance, it is  also reasonable to
assume that if a transition from state $\alpha$ to state $\beta$
occurs, then the reverse transition from $\beta$ to $\alpha$ occurs
with the same rate. This implies the condition
\begin{equation}
\lambda_\alpha \, A_{\beta,\alpha} = \lambda_\beta \, A_{\alpha, \beta} \; ,
\;\;\;  \alpha,\beta =1,\dots,N
\label{eq5_5}
\end{equation} 
If eq. (\ref{eq5_5}) holds, the GPK process is said to be
{\em transitionally symmetric}.
Unless otherwise stated, we assume that condition (\ref{eq5_5})
holds, which implies that the matrix ${\bf K}=(K_{\alpha,\beta})_{\alpha,\beta=1}^N$, $K_{\alpha,\beta}=\lambda_\beta \, A_{\alpha,\beta}$
is symmetric. Indicating with $\widehat{\boldsymbol \Lambda}=\mbox{diag}(\lambda_1,\dots,\lambda_N)$, the diagonal $N \times N$ matrix, the entries of 
which are the transition rates defining the vector ${\boldsymbol \Lambda}$,
eq. (\ref{eq5_5}) implies that 
\begin{equation}
{\bf K}= {\bf A} \, \widehat{\boldsymbol \Lambda} 
\label{eq5_6}
\end{equation}
is symmetric.

If all the transition rates coincide and thus are equal to a constant 
$\lambda$, the detailed balance condition (\ref{eq5_5}) implies that the
transition matrix ${\bf A}$ is symmetric,  thus, because of
eq. (\ref{eq5_2}), it is a doubly stochastic matrix.

Let us consider the velocity vectors $\{ {\bf b}_\alpha\}_{\alpha=1}^N$.
As the Poissonian perturbation represents a form of
unbiased stochastic motion, some conditions should be imposed
on these vectors. The nature of these conditions depends, in turn,
on the Markovian recombination mechanism amongst the $N$ distinct
states of $\chi_N(t)$.
Consider  the Markovian process (\ref{eq5_3}) for the
partial probabilities $P_\alpha(t)$, in the general case,
i.e., without setting any conditions on ${\bf A}$ other than
it is a left-stochastic matrix. These partial probabilities
converge for large time to the equilibrium values 
${\bf P}^*=(P_1^*,\dots,P_n^*)$ given by
\begin{equation}
\widehat{\boldsymbol \Lambda} \, {\bf P}^* = {\bf A} \, 
\widehat{\boldsymbol \Lambda} \, {\bf P}^*
\label{eq5_a1}
\end{equation}
Setting 
${\bf P}^*=\widehat{\boldsymbol \Lambda}^{-1} \, {\bf Q}^*$, eq. (\ref{eq5_a1})
becomes 
\begin{equation}
{\bf A} \, {\bf Q}^* = {\bf Q}^*
\label{eq5_a2}
\end{equation}
i.e., ${\bf Q}^*$ corresponds to the right Frobenius eigenvector ${\bf r}^{(1)}=(r_1^{(1)},\dots,r_N^{(1)})$, $r_\alpha^{(1)} \geq 0$,
of the left-stochastic matrix ${\bf A}$, associated
with the eigenvalue $1$, normalized in a suitable probabilistic way.
Therefore,
\begin{equation}
P_\alpha^* = \frac{ r_\alpha^{(1)}}{\lambda_\alpha}  \qquad
\alpha=1,\dots,N \,, \qquad \sum_{\alpha=1}^N \frac{r_\alpha^{(1)}}{\lambda_\alpha} =1
\label{eq5_a3}
\end{equation}
where the latter condition for ${\bf r}^{(1)}$ provides
the correct probabilistic normalization of ${\bf P}^*$.

As regards the velocity vectors ${\bf b}_\alpha$ forming ${\mathcal B}_N$,
the preliminary condition that they span ${\mathbb R}^n$
should be set  in order  to avoid degeneracy along some directions.
More precisely,
there is a subset ${\mathcal B}_{N^,n}^\prime \subseteq {\mathcal B}_N$ of
$n$ velocity vectors  forming a base for ${\mathbb R}^n$.
This automatically implies that $N \geq n$.
Moreover, a condition should be further imposed
in order to ensure that the stochastic perturbation
does not provide any finite mean drift, that is an {\em unbiasing condition}.
A suitable condition for the velocity vectors $\{{\bf b}_\alpha \}_{\alpha=1}^N$
ensuring unbiasing conditions is that with respect to
equilibrium distribution ${\bf P}^*$, the resulting stochastic
flux ${\bf J}_d^*= \sum_{\alpha=1}^N P_\alpha^* \, {\bf b}_\alpha $
vanishes. Therefore, we can assume
\begin{equation}
\sum_{\alpha=1}^N  P_\alpha^* \, {\bf b}_\alpha =
\sum_{\alpha=1}^N \frac{ r_\alpha^{(1)} \, {\bf b}_\alpha}{\lambda_\alpha}
=0
\label{eq5_a4}
\end{equation}
If the stochastic perturbation associated with the $N$-state finite
Poisson process is assumed in equilibrium conditions
starting from $t=0$, it follows that the initial
condition for $\chi_N(t)$ would be $\mbox{Prob}[\chi_N(0)=\alpha]= r_\alpha^{(1)}/\lambda_\alpha$, $\alpha=1,\dots,N$.

If all the transition rates are equal, $\lambda_\alpha=\lambda$,
$\alpha=1,\dots,N$, $A_{\alpha,\beta}$ is symmetric,
i.e., it is a doubly-stochastic matrix. The normalized
Frobenius eigenvalue is  associated with a uniform eigenvector
$r_\alpha^{(1)}=\lambda/ N$, $\alpha=1,\dots,N$, so that eq. (\ref{eq5_a4})
reduces to
\begin{equation}
\sum_{\alpha=1}^N {\bf b}_\alpha =0
\label{eq5_a5}
\end{equation}
We will see in paragraph \ref{sec_4_2} that eq. (\ref{eq5_a4})
reduces to eq. (\ref{eq5_a5}) whenever the matrix ${\bf A} \, \widehat{\boldsymbol \Lambda}$ is symmetric.

A GPK is {\em velocity-symmetric} if for any vector ${\bf b}_{\alpha} \in {\mathcal B}_N$,
there exists an $\alpha^*=1,\dots, N$, such that
\begin{equation}
{\bf b}_{\alpha^*}= - {\bf b}_{\alpha}
\label{eq5_7}
\end{equation}
For velocity-symmetric GPK, $N$ should be an even integer. 

Let us indicate with $b^{(c)}$ and $\lambda^{(c)}$
the characteristic values
\begin{equation}
b^{(c)}= \min_{1 \leq \alpha \leq n} |{\bf b}_\alpha| \; ,
\qquad \lambda^{(c)}= \min_{1 \leq \alpha \leq n} \lambda_\alpha
\label{eq5_9}
\end{equation}
where $| \cdot |$ indicates the Euclidean norm of a vector.
$b^{(c)}>0$ corresponds to the minimum velocity in the ${\mathcal B}_N$
system, and $\lambda^{(c)}>0$ to the slowest transition rate
in the recombination process amongst
the states. With these characteristic
quantities, the {\em nominal diffusivity} $D_{\rm nom}$ of a GPK
process can be defined as
\begin{equation}
D_{\rm nom}= \frac{(b^{(c)})^2}{2 \, \lambda^{(c)} }
\label{eq5_10}
\end{equation}
which plays a role in the assessment of the Kac limit.
Let us consider a GPK process in a domain characterized by a 
``characteristic length'' $L_c$ (i.e., by a characteristic value
of the state variable ${\bf x}$). A dimensionless group ${\mathcal K}$, referred
to as the {\em Kac number} can be defined as
\begin{equation}
{\mathcal K}= \frac{b^{(c)}}{ L_c \, \lambda^{(c)}} = \frac{ 2 \, D_{\rm nom}}
{L_c \, b^{(c)}}
\label{eq5_11}
\end{equation}
The Kac number is the ratio between the slowest recombination time
amongst the states of the process $t_{\rm recomb}=1/\lambda^{(c)}$
and the slowest characteristic advection time $t_{\rm adv}=L_c/b^{(c)}$
of the system of velocity vectors characterizing the stochastic
perturbation acting in a GPK process. 

It follows straightforwardly that all the higher-dimensional
extensions discussed in the previous Section can be
regarded as GPK processes. To give an example
consider the two-dimensional model eqs. (\ref{eq4_4})-(\ref{eq4_7})
 discussed in Section \ref{sec_4}.
In this case, all the
$\lambda_\alpha$, $\alpha=1,\dots,4$, coincide and are equal to $2 \, \lambda$.
The vectors ${\bf b}_\alpha$ are given by:
\begin{equation}
{\bf b}_{\alpha}= \sqrt{2} \, b \,
\left (
\begin{array}{c}
\cos(2 \pi (\alpha-1)/N+\pi/4) \\
\sin(2 \pi (\alpha-1)/N+\pi/4)
\end{array}
\right ) \; , \;\;\; \alpha=1,\dots,4
\label{eq5_11_a}
\end{equation}
 where state $\alpha=1$
 corresponds to $(-1)^{\chi_1(t)}=1$,
$(-1)^{\chi_2(t)}=1$, state $\alpha=2$ to  $(-1)^{\chi_1(t)}=-1$
, $(-1)^{\chi_2(t)}=1$, state $\alpha=3$ to $(-1)^{\chi_1(t)}=-1$
, $(-1)^{\chi_2(t)}=-1$, state  $\alpha=4$ to
$(-1)^{\chi_1(t)}=1$, $(-1)^{\chi_2(t)}=-1$.

\begin{figure}[h]
\begin{center}
{\includegraphics[height=4.5cm]{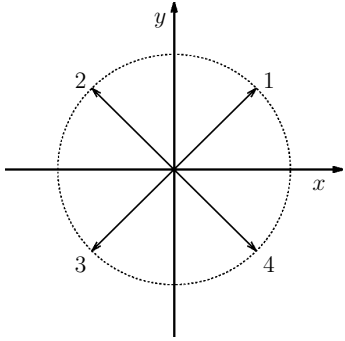}}
\end{center}
\caption{Structure  of states and velocity vectors for the
two-dimensional GPK process considered in the main text, eq. (\ref{eq5_11_a}).}
\label{Fig1}
\end{figure}

 The transition probability
matrix ${\bf A}$ is given by
\begin{equation}
{\bf A}= \left (
\begin{array}{llll}
0 & 1/2 & 0 & 1/2 \\
1/2 & 0 & 1/2 & 0 \\
0 & 1/2 & 0 & 1/2 \\
1/2 & 0 & 1/2 & 0
\end{array}
\right ) 
\label{eq5_11_b}
\end{equation}
Using the terminology introduced for GPK processes,
it is a velocity-symmetric process, possessing
a doubly stochastic symmetric matrix ${\bf A}$.

\subsection{Statistical description of GPK processes}
\label{sec_4_1}

Consider a GPK process $(n,N,{\boldsymbol \Lambda}, {\bf A},{\mathcal B}_N, {\bf v}({\bf x}))$, and let $\widehat{p}_{\alpha,\beta}({\bf x}, t /{\bf x}_0, t_0)$, $\alpha,\beta=1,N$  be   the transitional probabilities of finding
${\bf X}(t) \in ({\bf x},{\bf x}+d{\bf x})$ at time $t$ given
that ${\bf X}(t_0)={\bf x}_0$. From the Markovian transition
structure amongst the $N$-state finite Poisson process $\chi_N(t)$,
it follows that $\widehat{p}_{\alpha,\beta}({\bf x}, t /{\bf x}_0, t_0)$
satisfy the  hyperbolic equations
\begin{eqnarray}
\partial_t \widehat{p}_{\alpha,\beta}({\bf x}, t /{\bf x}_0, t_0)
& =  & - \nabla \cdot \left [
({\bf v}({\bf x})+ {\bf b}_\alpha) \,
\widehat{p}_{\alpha,\beta}({\bf x}, t /{\bf x}_0, t_0) \right ]
\label{eq5_12} \\
& -& \lambda_\alpha \widehat{p}_{\alpha,\beta}({\bf x}, t /{\bf x}_0, t_0)
+ \sum_{\gamma=1}^N \lambda_\gamma \, A_{\alpha,\gamma}
\, \widehat{p}_{\gamma,\beta}({\bf x}, t /{\bf x}_0, t_0)
\nonumber
\end{eqnarray}
$\alpha,\beta=1,\dots,N$.
Defining the partial probabilities $p_{\alpha}({\bf x},t)$
as the sum of $\widehat{p}_{\alpha,\beta}({\bf x}, t /{\bf x}_0, t_0)$
with respect to the initial state $\beta$ (and, for notational
simplicity neglecting the functional dependence on the initial
state), these quantities satisfy the
same balance equation (\ref{eq5_12}), namely
\begin{eqnarray}
\hspace{-1.5cm} \partial_t p_{\alpha}({\bf x}, t)
 =   - \nabla \cdot \left [
({\bf v}({\bf x})+ {\bf b}_\alpha) \,
p_{\alpha}({\bf x}, t) \right ]
& -& \lambda_\alpha p_{\alpha}({\bf x}, t)
+ \sum_{\gamma=1}^N \lambda_\gamma \, A_{\alpha,\gamma}
\, p_{\gamma}({\bf x}, t)
\label{eq5_13} 
\end{eqnarray}
$\alpha=1,\dots,N$. 
The quantities $\{ p_{\alpha}({\bf x},t)\}_{\alpha=1}^N$ represent
the primitive statistical variables of a GPK process,
while  the overall
probability density function $p({\bf x},t)$ and the
diffusive flux ${\bf J}_d({\bf x},t)$, i.e., the
flux associated with the stochastic perturbation,
are derived statistical quantities,
\begin{equation}
p({\bf x},t) = \sum_{\alpha=1}^N p_{\alpha}({\bf x},t)\; , \; \;\; 
{\bf J}_d({\bf x},t) = \sum_{\alpha=1}^N {\bf b}_\alpha({\bf x},t) \,
p_\alpha({\bf x},t)
\label{eq5_14}
\end{equation}

From eq. (\ref{eq5_13}) - or equivalently from the
underlying Markovian nature of the $N$-state
finite Poisson process generating stochasticity in the
system - it follows that a GPK process 
satisfies the extended Markov property,
\begin{equation}
\widehat{p}_{\alpha,\beta}({\bf x},t / {\bf x}_0,t_0) =
\sum_{\gamma=1}^N \int 
\widehat{p}_{\alpha,\gamma}({\bf x},t / {\bf y}_0,t_1)
\, 
\widehat{p}_{\gamma,\beta}({\bf y},t_1 / {\bf x}_0,t_0) \, d {\bf y}
\label{eq5_15}
\end{equation}
for any $t_0 < t_1 < t$, $\alpha,\beta=1,\dots,N$. 
Starting from this extended Markovian
condition the adjoint formalism of GPK process can be derived
\cite{giona_adjoint}.

\subsection{Examples and further observations}
\label{sec_4_2}

This paragraph discusses several examples in order  to  highlight some
 properties and additional 
observations related to GPK processes.

\begin{figure}[h]
\begin{center}
{\includegraphics[height=7cm]{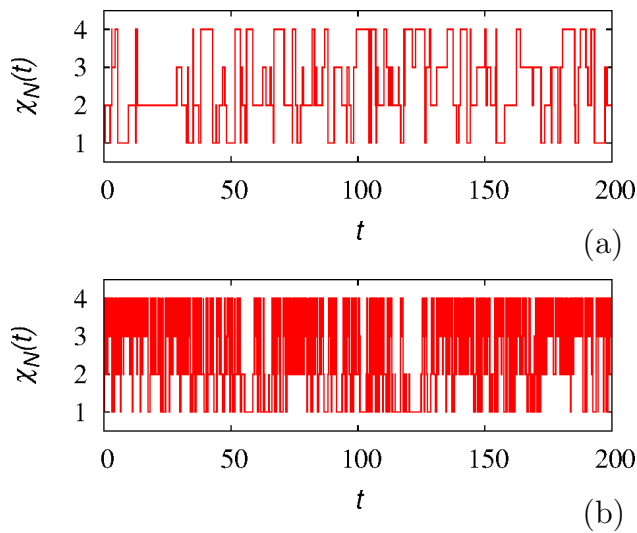}}
\end{center}
\caption{Realizations of a $N$-state finite Poisson process with
$N=4$. Panel (a) uniform transition rates $\lambda_\alpha=1$,
$A_{\alpha,\beta}=1/N$. Panel (b) process characterized by
non-uniform transition rates associated with the
symmetric matrix ${\bf K}$ given by eq. (\ref{eq5_b5}).}
\label{Fig2}
\end{figure}

Figure \ref{Fig2} panel (a) depicts a realization of
a $N$-state finite Poisson process $\chi_N(t)$ with
$N=4$, $\lambda=1$, using the symmetric transition
matrix $A_{\alpha,\beta}=1/N$, $\alpha,\beta=1,\dots,N$.

Making use of this process, and introducing the
system of velocity vectors ${\bf b}_\alpha=(\cos(2 \pi (\alpha-1)/N), 
\sin(2 \pi(\alpha-1)/N))$, $\alpha=1,\dots,N$, a GPK process in ${\mathbb R}^2$
 can be defined
in the absence of a deterministic bias, i.e., ${\bf v}({\bf x})=0$.

\begin{figure}[h]
\begin{center}
{\includegraphics[height=4.5cm]{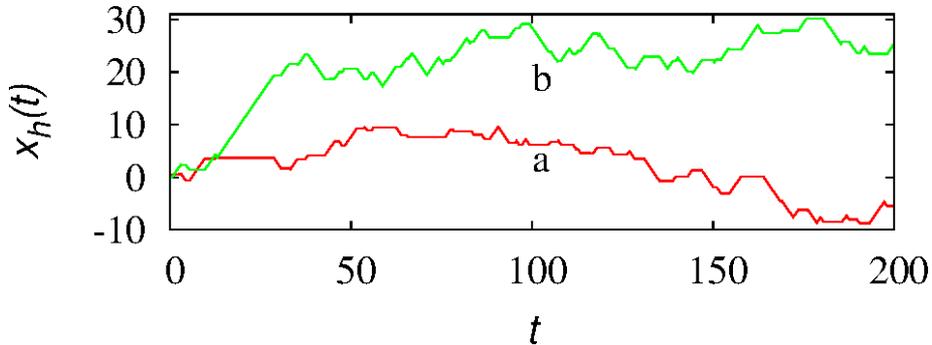}}
\end{center}
\caption{Realization of a two-dimensional GPK process associated
with $\chi_N(t)$ depicted in figure \ref{Fig2} panel (a). Line (a)
refers to $x_1(t)$, line (b) to $x_2(t)$.}
\label{Fig3}
\end{figure}

Figure \ref{Fig3} shows a portion of the trajectory ${\bf x}(t)=(x_1(t),x_2(t))$
of the GPK process defined above starting from ${\bf x}(0)=0$.

Although it has been already mentioned, it is worth
pointing out that the trajectories of a GPK process are
almost everywhere smooth functions of time $t$.
This means that ${\bf x}(t)$ is $C^\infty$ for almost all $t>0$
apart from a numerable set of time instants $t_j^*$, $j=1,2,\dots$,
where the derivative of ${\bf x}(t)$ is not defined. These
discontinuities correspond to the  transition instants
of the $N$-state finite Poisson process from a state $\alpha=\chi_N(t_{h,-}^*)$,
to a state $\beta=\chi_N(t_{h,+}^*)$, $t_{h,\pm}^*=\lim_{\varepsilon \rightarrow
0} t_h^*\pm \varepsilon$. Correspondingly,  the
left- and right-derivatives of ${\bf x}(t)$
at time $t_h^*$
are still defined and
\begin{equation}
\left . \frac{d {\bf x}(t)}{d t} \right |_{t=t_{h,+}^*}
- \left . \frac{d {\bf x}(t)}{d t} \right |_{t=t_{h,-}^*}
= {\bf b}_{\chi_N(t_{h,+}^*)}- {\bf b}_{\chi_N(t_{h,-}^*)}
\label{eq5_b1}
\end{equation}

Figure \ref{Fig2} panel (b) depicts the realization
of a $N$-state finite Poisson process, $N=4$, in the
presence of a nonuniform distribution of the
transition rates $\lambda_\alpha$, $\alpha=1,\dots,N$.

Assume that the detailed balance condition (\ref{eq5_5}) hold,
i.e., consider a transitionally symmetric GPK process.
For this class of GPK processes, the symmetric matrix ${\bf K}$
defined by  eq. (\ref{eq5_6}) fully characterizes the transition
properties of the process.  By definition
\begin{equation}
K_{\alpha,\beta}= K_{\beta,\alpha}\, , \;\; \; K_{\alpha,\beta} \geq 0
\, ,   \;\;\; \sum_{\alpha=1}^N K_{\alpha,\beta} >0 
\label{eq5_b2}
\end{equation}
From the structure of ${\bf K}$, the vector ${\boldsymbol \Lambda}$
of the transition rate
and the left-stochastic matrix ${\bf A}$ of the transition
probabilities can be determined as
\begin{equation}
\lambda_{\alpha}=\sum_{\beta=1}^N K_{\beta,\alpha} >0 \, , \qquad
\alpha=1,\dots,N
\label{eq5_b3}
\end{equation}
corresponding to the sum of entries of the $\alpha$-th column
of ${\bf K}$, and
\begin{equation}
A_{\alpha,\beta}= \frac{K_{\alpha,\beta}}{\lambda_\beta} \, ,
\qquad \alpha,\beta=1,\dots,N
\label{eq5_b4}
\end{equation}
This provides an alternative way of defining a transitionally
symmetric GPK process
starting from the
quintuple
\[
(n,N,{\bf K},
 {\mathcal B}_N,{\bf v}({\bf x}))
\]
where ${\bf K}$ satisfies  conditions (\ref{eq5_b2}), out
of which ${\boldsymbol \Lambda}$ and ${\bf A}$ can be
derived applying eqs. (\ref{eq5_b3})-(\ref{eq5_b4}).

For example, the non-uniform $N$-state finite Poisson process,
a realization of which is depicted in figure \ref{Fig2} panel (b),
has been obtained starting from the symmetric non-negative matrix
\begin{equation}
{\bf K}=
\left (
\begin{array}{cccc}
0 & 1 & 0.1 & 0.8 \\
1 & 0 & 0.2 & 2 \\
0.1 & 0.2 & 0 & 0.3 \\
0.8 & 2 & 0.3 & 0
\end{array}
\right )
\label{eq5_b5}
\end{equation}
to which the transition-rate vector ${\boldsymbol \Lambda}=(1.9,3.2,0.6,3.1)$
is associated with.

While the use of the ${\bf K}$-representation is more
convenient for analytical calculations, the representation
via ${\bf A}$ and ${\boldsymbol \Lambda}$ is
more suitable for  stochastic simulations of
the microdynamic equation of motion  (\ref{eq5_4}).
Let us explore further the implications of this observation.

Consider a transitionally  symmetric GPK processes
defined by a symmetric non-negative matrix ${\bf K}$.
The stationary probabilities for the $N$-state finite
Poisson process generating it are given by
\begin{equation}
P_\alpha^*= \frac{1}{N} \, , \;\;\; \alpha=1,\dots,N
\label{eq5_b6}
\end{equation}
for any choice of $\lambda_\alpha>0$.
This property follows immediately from  the fact
that
\begin{equation}
-\lambda_\alpha \, \frac{1}{N} + \sum_{\gamma=1}^N K_{\alpha,\gamma}
\, \frac{1}{N} = 
-\frac{1}{N} \sum_{\gamma=1}^N \left (K_{\gamma,\alpha} - K_{\alpha,\gamma}
\right )  =  0
\label{eq5_b7}
\end{equation}
due to the  symmetry of $K_{\alpha,\gamma}$. As  byproduct of it,
the zero-bias condition for a generic transitionally symmetric GPK
process reduces to  eq. (\ref{eq5_a5}).
From eq. (\ref{eq5_b6}) and from the definition
of the vector ${\bf Q}^*$
entering eq.  (\ref{eq5_a2}), it follows that
\begin{equation}
Q_\alpha^* = C \, \lambda_\alpha \, , \; \; \; \alpha=1,\dots,N \, ,
\;\;\; C = \left ( \sum_{\alpha=1}^N \lambda_\alpha \right )^{-1}
\label{eq5_b8}
\end{equation}
which implies that the entries of the right Frobenius eigenvector 
of the transition probability matrix ${\bf A}$ are proportional
to the transition rates. This property is useful in order to address the
statistics of the transition times (see below).

\begin{figure}[h]
\begin{center}
{\includegraphics[height=6cm]{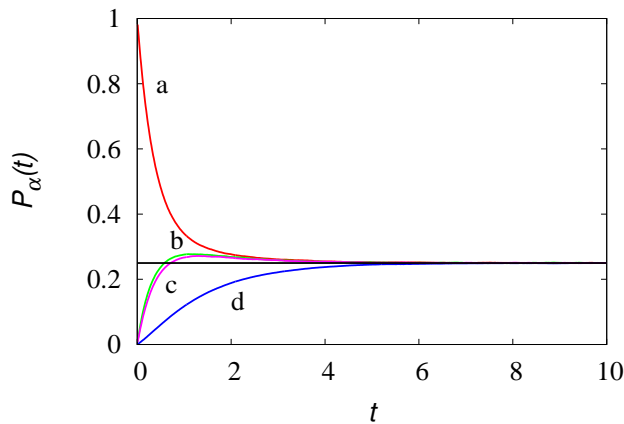}}
\end{center}
\caption{Probabilities $P_\alpha(t)$, $\alpha=1,\dots,4$,
for the $4$-state finite Poisson process defined by
the ${\bf K}$ matrix (\ref{eq5_b5}) obtained
from stochastic simulations of the process
starting from $P_\alpha(0)=\delta_{\alpha,1}$. From
line (a) to line (d), $\alpha=1,2,3,4$. The horizontal
line represents $P_\alpha^*=1/4$.}
\label{Fig4}
\end{figure}

Conversely, the numerical  simulation of eq. (\ref{eq5_4})
is simply based on the Markovian dynamics of $\chi_N(t)$,
and can be  conveniently built up
 using the $({\bf A},{\boldsymbol \Lambda})$-representation. A stochastic algorithm for simulating eq. (\ref{eq5_4})
can be organized as follows:
\begin{enumerate}
\item {\it Let ${\bf x}(t_0)={\bf x}_0$, $\chi_N(t_0)=\alpha_0$ be the initial
condition at time $t_0$;}
\item {\it Pick   a transition time $\tau$ out of the
exponential distribution $p_{\tau}(\tau / \alpha_0) = \lambda_{\alpha_0}
\, e^{-\lambda_{\alpha_0} \tau}$ pertaining to the $\alpha_0$-th state;}
\item {\it For times $t<t_0+\tau$ integrate the equation
of motion $d {\bf x}(t)/dt = {\bf v}({\bf x}(t)) + {\bf b}_{\alpha_0}$;}
\item {\it At time $t^*_1=t_0 +\tau$ the state-transition occurs: select 
a new state $\beta$ chosen randomly according to
the transitional probabilities $\pi_\beta=A_{\beta,\alpha_0}$,
$\beta=1,\dots,N$;}
\item {\it Continue the procedure using ${\bf x}(t_0+\tau)$ and $\beta$
as the new initial condition.}  
\end{enumerate}
It follows from the sketch of the numerical algorithm outlined
above that the decomposition of ${\bf K}$ into a left-stochastic
matrix ${\bf A}$ and into a vector of transition rate ${\boldsymbol \Lambda}$
is functional to the integration of eq. (\ref{eq5_4}), in
which the central core of the algorithm is the simulation of the stochastic
process $\chi_N(t)$.
To give a numerical example, figure \ref{Fig4} depicts 
the evolution of the probabilities $P_\alpha(t)$ associated
with the states of $\chi_N(t)$ for the
model described by the symmetric ${\bf K}$ matrix eq. (\ref{eq5_b5}),
starting from $P_{\alpha}(0)=\delta_{\alpha,1}$.
Data refer to an ensemble of $10^6$ particles.
As expected from eq. (\ref{eq5_b6}), $P_{\alpha}(t) \rightarrow P_\alpha^*=1/4$,
$\alpha=1,\dots,4$ for large times.

A more delicate issue is related to the distribution of the
switching times $\tau$ in  GPK processes. Let
$\tau_j=t_{j+1}^*-t_j^*$, $j=1,2,\dots$, be the sequence of
intervals between two consecutive transitions of $\chi_N(t)$,
and $p_\tau(\tau)$ its probability density function. It follows
from the way the stochastic dynamics is generated, that $p_\tau(\tau)$
is the average of the probability density function of the switching
intervals
$p_\tau(\tau/\alpha)=\lambda_\alpha \, e^{-\lambda_\alpha \tau}$
starting from the state $\alpha$ weighted with respect to
the stationary probability distribution ${\bf P}_{S,\alpha}^*=(P_{S,1}^*,\dots,P_{S,N}^*)$ of finding the state $\alpha$ just at the switching times $t_j^*$,
\begin{equation}
p_\tau(\tau)= \sum_{\alpha=1}^N p_\tau(\tau/\alpha) \, P_{S,\alpha}^*
\label{eq5_b9}
\end{equation}

\begin{figure}[h]
\begin{center}
{\includegraphics[height=4.5cm]{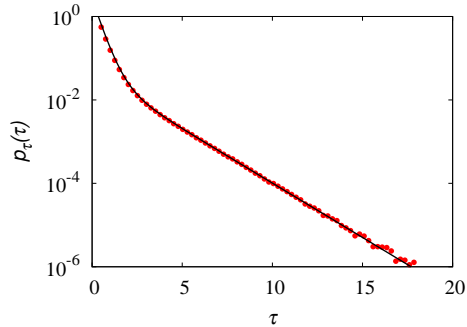}}
\end{center}
\caption{Probability density function for the
time interval between two consecutive transitions $p_\tau(\tau)$
vs $\tau$ for the $4$-state process defined
by the symmetric ${\bf K}$ matrix (\ref{eq5_b5}).
Solid line refers to eq. (\ref{eq5_b10}), symbols ($\bullet$)
to stochastic simulations.}
\label{Fig5}
\end{figure}

\begin{figure}[h]
\begin{center}
{\includegraphics[height=4.5cm]{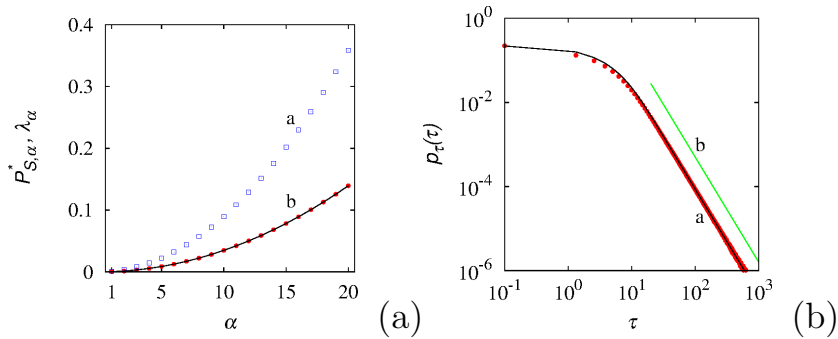}}
\end{center}
\caption{Statistical properties of the $N$ state process generated
by the matrix  eq. (\ref{eq5_b11}) for $N=20$. Panel (a):
distribution of transition rates $\lambda_\alpha$ vs $\alpha$
(symbols $(\square)$, indicated with the label (a)), and
probability distribution at the transition points $P_{S,\alpha}^*$
(line b and symbols ($\bullet$)).
Line (b) refers to the theoretical prediction (\ref{eq5_b8}),
symbols ($\bullet$) to stochastic simulation data.
Panel (b): probability density function $p_\tau(\tau)$
of time intervals between two consecutive transitions.
Symbols ($\bullet$) refers to stochastic simulations,
line (a) to the theoretical expression (\ref{eq5_b10}).
Line (b) indicates the power-law scaling $p_\tau(\tau) \sim \tau^{-5/2}$.}
\label{Fig6}
\end{figure}

Since the transition amongst the states of the process $\chi_N(t)$
is controlled exclusively by the transition probability matrix
${\bf A}$, it follows that ${\bf P}_{S}^*$ equals the
normalized right Frobenius eigenvector of ${\bf A}$, i.e.,
${\bf P}^*_S={\bf Q}^*$. Therefore, using eq. (\ref{eq5_b8})
one finally obtains for $p_\tau(\tau)$ the expression
\begin{equation}
p_\tau(\tau) = \frac{\sum_{\alpha=1}^N \lambda_\alpha^2 \, e^{-\lambda_\alpha
\, \tau}}{\sum_{\alpha=1}^N \lambda_\alpha}
\label{eq5_b10}
\end{equation}
Figure \ref{Fig5} depicts the comparison of eq. (\ref{eq5_b10})
against stochastic simulations for the process described
by the ${\bf K}$ matrix eq. (\ref{eq5_b5}). Simulation
data have been obtained using a sample of $N_S=10^8$
intervals between two consecutive transitions. The agreement is excellent.
A more severe check of eq. (\ref{eq5_b10}) can be obtained
by considering the model described by the symmetric matrix
\begin{equation}
K_{\alpha,\beta}= \frac{\alpha^2 \, \beta^2}{N^5} \, ,
\;\;\; \alpha,\beta=1,\dots,N
\label{eq5_b11}
\end{equation}
for a large number of states $N$.
Figure \ref{Fig6} (symbol $\square$) depicts the values
of $\lambda_\alpha$ obtained from this matrix for $N=20$.
As can be observed, the switching rates $\lambda_\alpha$ are distributed
almost uniformly in the interval $[0,\lambda_{\rm max}]$,
where $\lambda_{\rm max} \simeq 0.36$. Symbols $(\bullet)$ and
line (b) in figure \ref{Fig6}
depict   the entries of the
state distribution probability vector
${\bf P_S}^*$ at the transition times. Line (b) refers
to the theoretical expression (\ref{eq5_b8}) (since ${\bf P}_S^*={\bf Q}^*$),
while symbols $(\bullet)$ are simulation results obtained from a sample of 
 $2 \times 10^8$
transitions. 
The behavior of $p_\tau(\tau)$ for this model is depicted
in panel (b). Also in this case the agreement is excellent.
It is worth observing that the structure of the transition
rates that are distributed almost uniformly close to $\lambda=0$
determines a very anomalous profile of $p_\tau(\tau)$,
that for more than two decades scales as a power law $p_\tau(\tau) \sim \tau^{-5/2}$ (line b in panel (b)).

\subsection{Boundary conditions in closed systems}
\label{sec_3_3}

In this article we analyze mainly the unbounded
propagation of GPK processes in ${\mathbb R}^n$,
so that the only condition to be imposed is the
regularity at infinity for the partial probability 
 waves $p_\alpha({\bf x},t)$.
Nevertheless, keeping the focus on probabilistic problems, it is
useful to address also the case of GPK process in bounded domain $\Omega$,
assuming probability conservation, namely
\begin{equation}
\frac{d}{d t} \int_\Omega \sum_{\alpha=1}^N p_\alpha({\bf x},t) \, d {\bf x}=0
\label{eq5_c1}
\end{equation}
Two cases are relevant. The first is represented by compact manifold,
such as the $n$-torus ${\mathcal T}^n=\{ {\bf x} \; | 0 \leq x_h < 1\, ,
\; h=1,\dots,n \}$, where $x_h=0$ and $x_h=1$ correspond to the
one and the same point for $h=1,\dots,n$. In this case, periodic boundary
conditions apply to the partial probability waves, namely
\begin{equation}
p_{\alpha}({\bf x},t) |_{x_h=0} = p_{\alpha}({\bf x},t) |_{x_h=1}
\; , \alpha=1,\dots,N\;\;\; h=1,\dots,n 
\label{eq5_c2}
\end{equation}

The second case is when $\Omega$ is a bounded set of ${\mathbb R}^n$,
possessing a boundary $\partial \Omega$. Let ${\bf n}_e({\bf x})$
be the normal unit vector pointing outwardly at ${\bf x} \in \partial \Omega$,
and suppose that the deterministic field ${\bf v}({\bf x})$
admit a vanishing normal component at $\partial \Omega$,
\begin{equation}
{\bf v}({\bf x}) \cdot {\bf n}_e({\bf x}) =0 \; , \;\;\; {\bf x} \in \partial
\Omega
\label{eq5_c3}
\end{equation}
The condition ensuring probability conservation in $\Omega$
can be exclusively expressed in terms of the diffusive
flux ${\bf J}_d$, namely ${\bf J}_d \cdot {\bf n}_e=0$ at $\partial \Omega$.
Expressed with respect to the partial probability waves, this condition becomes
\begin{equation}
\sum_{\alpha=1}^N p_{\alpha}({\bf x},t) \left [ {\bf b}_\alpha \cdot
{\bf n}_e({\bf x}) \right ] =0 \, , \;\;\; {\bf x} \in \partial \Omega
\label{eq5_c4}
\end{equation}
Set ${\bf b}_\alpha \cdot {\bf n}_e({\bf x})= |{\bf b}_\alpha|
\, \cos_{n,\alpha}({\bf x})$. Upon a suitable relabeling of the states,
the
partial waves at ${\bf x} \in \partial \Omega$  can be subdivided
into three subsets: (i) probability waves
that propagate outwardly $\Omega$, $\alpha^\prime=1,\dots,N_{\rm out}$
such that $\cos_{n,\alpha^\prime}({\bf x})>0$, (ii) probability waves,
$\alpha^{\prime \prime}=N_{\rm out}+1,\dots,N_{\rm out}+N_{\rm in}$
propagating inwardly, i.e,  $\cos_{n,\alpha^{\prime \prime}}({\bf x})<0$,
and (iii) neutral waves, $\alpha^{\prime \prime \prime}=N_{\rm out}+N_{\rm in}+1,\dots, N$, such that their direction is orthogonal to
${\bf n}_e({\bf x})$, i.e., $\cos_{n,\alpha^{\prime \prime \prime}}({\bf x})=0$

Using this decomposition, eq. (\ref{eq5_c4}) can be formulated as:
\begin{equation}
\sum_{\alpha^{\prime \prime}=N_{\rm out}+1}^{N_{\rm out}+N_{\rm in}}
p_{\alpha^{\prime \prime}}({\bf x},t) \, |{\bf b}_{\alpha^{\prime \prime}}|
\, |\cos_{n,\alpha^{\prime \prime}}({\bf x})| =
\sum_{\alpha^\prime=1}^{N_{\rm out}} 
p_{\alpha^{\prime}}({\bf x},t) \, |{\bf b}_{\alpha^\prime}|
\, |\cos_{n,\alpha^\prime}({\bf x})|
\label{eq5_c5}
\end{equation}
expressing the property that the inwardly oriented flux
associated with the back-scattered partial waves $p_{\alpha^{\prime \prime}}({\bf x},t)$ at $\partial \Omega$ should compensate the outwardly oriented
flux associated with $p_{\alpha^\prime}({\bf x},t)$.

The zero-flux boundary condition expressed by eq. (\ref{eq5_c5}) can
be realized in many different ways. The simplest choice is to
consider an equipartition procedure, meaning that we
assume that all the back-scattered partial probabilities $p_{\alpha^{\prime
\prime}}({\bf x},t)$ are equal to each other, so that
eq. (\ref{eq5_c5}) implies
\begin{equation}
p_{\alpha^{\prime \prime}}({\bf x},t) =
\frac{ \sum_{\alpha^\prime=1}^{N_{\rm out}} 
p_{\alpha^{\prime}}({\bf x},t) \, |{\bf b}_{\alpha^\prime}|
\, |\cos_{n,\alpha^\prime}({\bf x})|}
{\sum_{\alpha^{\prime \prime}=N_{\rm out}+1}^{N_{\rm out}+N_{\rm in}}
 |{\bf b}_{\alpha^{\prime \prime}}|
\, |\cos_{n,\alpha^{\prime \prime}}({\bf x})|}
\label{eq5_c6}
\end{equation}
$\alpha^{\prime \prime} =N_{\rm out}+1,\dots,N_{out}+N_{\rm in}$.

In terms of the stochastic microdynamics (\ref{eq5_4}), 
the impermeability condition at $\partial \Omega$, does
not imply solely a reflecting boundary condition for ${\bf x}(t)$
at any point of $\partial \Omega$, as for Langevin dynamics
driven by Wiener fluctuations, but also a transition
to  a new state, induced 
 by
the reflecting boundary itself, and not by the internal statistical 
structure
of the $N$-state finite Poisson process $\chi_N(t)$ \cite{giona_bc}.
If ${\bf x}(t)$ reaches the
boundary at time $t$, this means that the state $\alpha$ of $\chi_N(t)$
is such that $\cos_{n,\alpha}({\bf x}(t))>0$, i.e., its stochastic velocity
vector is pointing outwardly. After the reflection, say at $t+\Delta t$,
the state of $\chi_N(t+\Delta t)$ also changes, taking any value
between $\alpha^{\prime \prime}=N_{\rm out}+1,\dots, N_{\rm out}+N_{\rm in}$,
corresponding to an inwardly oriented direction. Assuming
the equipartition rule expressed by eq. (\ref{eq5_c6}), it corresponds
to the  equiprobable 
 choice of any of the $N_{\rm in}$ inwardly oriented states.

\section{The Kac limit}
\label{sec_5}

The principle of asymptotic Kac convergence is, in some sense,
a closure condition with respect to  the  Brownian motion
paradigm. In the limit for $b^{(c)}$
and $\lambda^{(c)}$ tending to infinity, keeping fixed the
nominal diffusivity $D_{\rm nom}$, a GPK process should converge to
a strictly Markovian process described solely by the overall probability
density function $p({\bf x},t)$, solution of a parabolic advection-diffusion
equation, where the advective term is controlled by the deterministic velocity
field ${\bf v}({\bf x})$, and the diffusive term is characterized
by an effective diffusivity $D_{\rm eff}$ proportional to $D_{\rm nom}$.
The relation between $D_{\rm eff}$ and $D_{\rm nom}$ depends on the structure
of the GPK process i.e., on ${\mathcal B}_N$, ${\bf A}$ and ${\boldsymbol 
\Lambda}$, as will be clear in the remainder.

The assessment of the Kac limit for  generic GPK processes is not
a simple mathematical issue, as discussed by Kolesnik for some, relatively
simple two-dimensional Poisson-Kac processes \cite{kolesnik3}. 
A physically-oriented strategy for tackling the Kac limit is
developed below, enforcing the fact that this limit corresponds to
an infinitely fast recombination kinetics  amongst the partial probability
waves, and this implies a form of thermalization/equipartition
of  probability amongst the partial waves.

The assessment of the Kac limit provides also the  functional
conditions on the structure of ${\mathcal B}_N$, ${\bf A}$ and ${\boldsymbol \Lambda}$, i.e., a system of linear constraints relating the 
values of the stochastic velocity vectors to
 the structure of the transition matrix
and to the values of the transition rates.

For generic GPK processes, the balance equation for the
overall probability density function $p({\bf x},t)$ can be
always written as
\begin{equation}
\partial_t p({\bf x},t)= - \nabla \cdot \left [ {\bf v}({\bf x}) 
\, p({\bf x},t) \right ]
- \nabla \cdot {\bf J}_d({\bf x},t)
\label{eq6_0}
\end{equation}
where ${\bf J}_d({\bf x},t)$ is the diffusive flux associated with the
stochastic perturbation. Therefore, the assessment of the
Kac limit involves the structure of the constitutive equation
for ${\bf J}_d({\bf x},t)$, and implies a limit  Fickian behavior
for ${\bf J}_d({\bf x},t) \sim \nabla p({\bf x},t)$. 

The analysis of the Kac limit is developed in several steps. In the
first paragraph we develop the criteria for the assessment of the
Kac limit for GPK processes of increasing complexity. 
Next, we present some numerical examples of the theory developed.
Section \ref{sec_6} is also related to the Kac limit,
from another perspective, namely its occurrence as an
emerging long-term property for finite values of $b^{(c)}$ and $\lambda^{(c)}$.

\subsection{The Kac limit for GPK process}
\label{sec_5_1}

As a first case, consider a GPK process in the
presence of a uniform distribution of transition
rates $\lambda_\alpha=\lambda$, $\alpha=1,\dots,N$, where
the transition matrix ${\bf A}$ is doubly stochastic (and {\em a fortiori} symmetric),
and assume that all the stochastic velocity vectors ${\bf b}_\alpha$
possess the same modulus,
\begin{equation}
{\bf b}_\alpha = b \, \widehat{\bf b}_\alpha \,, \qquad  |\widehat{\bf b}_\alpha|=1 \,, \qquad \alpha=1,\dots,N
\label{eq6_2}
\end{equation}
so that $\lambda^{(c)}=\lambda$, $b^{(c)}=b$.
The partial probability  waves $p_\alpha({\bf x},t)$ evolve
according to eq. (\ref{eq5_13}) where  all the $\lambda_\alpha$'s equal 
$\lambda$.
Multiplying eq. (\ref{eq5_13}) by ${\bf b}_\alpha$, and summing over the
states $\alpha$, the following constitutive equation for the
diffusive flux is obtained
\begin{eqnarray}
\partial_t {\bf J}_d({\bf x},t) & = &  - \nabla \cdot \left [ {\bf v}({\bf x}) \, {\bf J}_d({\bf x},t)
\right ] - \nabla \cdot \left  [ \sum_{\alpha=1}^N {\bf b}_\alpha \, 
{\bf b}_\alpha \, p_\alpha({\bf x},t) \right ] \nonumber \\
& - & \lambda \, {\bf J}_d({\bf x},t) + \lambda \, \sum_{\alpha,\gamma=1}^N
{\bf b}_\alpha \, A_{\alpha,\gamma} \, p_\gamma({\bf x},t)
\label{eq6_3}
\end{eqnarray}
In eq. (\ref{eq6_3}) we have used the notation ${\bf v}({\bf x}) \, {\bf J}_d({\bf x},t)$
and $\sum_{\alpha=1}^N {\bf b}_\alpha \, {\bf b}_\alpha$ to indicate
dyadic tensors, and $\nabla \cdot \left [ {\bf u}({\bf x}) \, {\bf w}({\bf x}) \right ]$ 
for the divergence of a tensorial quantity.
Given a dyadic tensor ${\bf u}({\bf x}) \, {\bf w}({\bf x})$,
where ${\bf v}({\bf x})=(v_1({\bf x}),\dots,v_n({\bf x}))$, ${\bf w}({\bf x})=(w_1({\bf x}),\dots,w_n({\bf x}))$,
the tensor divergence $\nabla \cdot \left [ {\bf u}({\bf x}) \, {\bf w}({\bf x}) \right]$
is defined componentwise as
\begin{equation}
\left \{\nabla \cdot \left [ {\bf u}({\bf x}) \, {\bf w}({\bf x}) \right ] \right  \}_h
= \sum_{h=1}^n \frac{\partial ( v_i \, w_h)}{\partial x_i} \,,  \qquad
h=1,\dots,n
\label{eq6_4}
\end{equation} 
where $\left \{ \nabla \cdot \left [ {\bf u}({\bf x}) \, {\bf w}({\bf x}) \right ] \right \}_h$ is the
$h$-th entry of $\nabla \cdot \left [ {\bf u}({\bf x}) \, {\bf w}({\bf x}) \right]$.

Let us assume that the stochastic velocity vectors belonging to ${\mathcal
B}_N$ satisfy the conditions
\begin{equation}
\sum_{\alpha=1}^N A_{\alpha,\gamma} \, {\bf b}_\alpha = \delta \,
{\bf b}_\gamma \, , \qquad \gamma=1,\dots,N
\label{eq6_5}
\end{equation}
where $\delta$ is some constant independent of $\gamma$.
Componentwise, this implies for any $\gamma$ 
\begin{equation}
\sum_{\alpha=1}^N A_{\alpha,\gamma} \, b_{\alpha,h} = 	
\delta \, b_{\gamma,h} \, , \qquad h=1,\dots,n
\label{eq6_6}
\end{equation}
This condition is apparently similar to an eigenvalue
equation, but it is not. It essentially expresses
a uniform transition condition for  the stochastic velocities  that
relates them
to the transition mechanism characterizing the GPK process,
described by the matrix $A_{\alpha,\gamma}$.
It is easy to check that eqs. (\ref{eq6_5}) are consistent with
the zero-bias condition (\ref{eq5_a5}). Summing over $\gamma,$ and enforcing the
doubly stochastic nature of $A_{\alpha,\gamma}$, one obtains
\begin{equation}
\sum_{\alpha=1}^N {\bf b}_\alpha = \delta \, \sum_{\gamma=1}^N
{\bf b}_{\gamma} 
\label{eq6_7}
\end{equation}
Therefore, either $\delta=1$, and the sum of the ${\bf b}_\alpha$
can attain any value, or $\delta \neq 1$ and $\sum_{\alpha=1}^N 
{\bf b}_\alpha=0$.
But the constant $\delta$ cannot be equal to $1$. This follows
from the following argument.
Taking the scalar product of both sides of eq. (\ref{eq6_6}) by
${\bf b}_\gamma$, 
\begin{equation}
\delta = \sum_{\alpha=1}^N A_{\alpha,\gamma} \, \widehat{\bf b}_\alpha \cdot 
\widehat{\bf b}_\gamma
\label{eq6_8}
\end{equation}
But, $|\widehat{\bf b}_\alpha \cdot \widehat{\bf b}_\gamma| \leq 1$,
and the equal sign occurs solely if $\alpha=\gamma$  or 
$\widehat{\bf b}_\gamma=-\widehat{\bf b}_\alpha$.
 Therefore,
\begin{equation}
|\delta | < \sum_{\alpha=1}^N A_{\alpha,\gamma} = 1
\label{eq6_9}
\end{equation}
It follows that, if eq. (\ref{eq6_6}) is fulfilled, than the zero-bias
condition  (\ref{eq5_a5}) is automatically satisfied. Moreover
$|\delta| <1$. Substituting this result into eq. (\ref{eq6_3}), one
obtains
\begin{eqnarray}
(1-\delta) \, {\bf J}_d({\bf x},t) & + & \frac{1}{\lambda} \left \{
\partial_t {\bf J}_d({\bf x},t) + \nabla \cdot \left [ 
{\bf v}({\bf x}) \, {\bf J}_d({\bf x},t) \right ] \right \} \nonumber \\
& = & - \frac{b^2}{\lambda} \,  \nabla \cdot \left [ \sum_{\alpha=1}^N
\widehat{\bf b}_\alpha \, \widehat{\bf b}_\alpha \, p_\alpha({\bf x},t)
 \right ] 
\label{eq6_10}
\end{eqnarray}
In the Kac limit, the transition rate $\lambda$ diverges, implying
that the recombination kinetics between the partial waves becomes
infinitely fast. It is reasonable therefore to
assume a principle of {\em local equilibrium}, namely that,
for very large $\lambda$,
\begin{equation}
p_\alpha({\bf x },t) = \frac{1}{N} \, p({\bf x},t) + o(\lambda^{-1})
\label{eq6_11}
\end{equation}
where $o(\lambda^{-1})$ is a quantity vanishing to zero for
$\lambda \rightarrow \infty$. Eq. (\ref{eq6_11}) represents
a form of thermalization of the internal state of the stochastic
process expressed via its partial probability  waves $p_\alpha({\bf x},t)$.
Enforcing eq. (\ref{eq6_11}),  the term at the right-hand
side of eq. (\ref{eq6_10}) becomes
\begin{equation}
 \nabla \cdot \left [ \sum_{\alpha=1}^N
\widehat{\bf b}_\alpha \, \widehat{\bf b}_\alpha \, p_\alpha({\bf x},t)
 \right ]
= \nabla \cdot \left [ {\bf S} \, p({\bf x},t) \right ]
 + o(\lambda^{-1})
\label{eq6_12}
\end{equation}
where 
\begin{equation}
 {\bf S}= \frac{1}{N} \sum_{\alpha=1}^N \widehat{\bf b}_\alpha
\, \widehat{\bf b}_\alpha
\label{eq6_12bis}
\end{equation}
represents a  symmetric dyadic tensor ${\bf S}$ referred to as the
{\em structure tensor} of the GPK process.
In the Kac limit, a GPK process reduces to a classical 
advection-diffusion equation characterized 
by a scalar diffusivity $D_{\rm eff}$
provided that the structure tensor is isotropic, i.e.,
\begin{equation}
S_{h,k}= \frac{1}{N}  \sum_{\alpha=1}^N \widehat{b}_{\alpha,h}
\, \widehat{b}_{\alpha,k}= \kappa \, \delta_{h,k}
\, , \qquad h,k=1,\dots, n
\label{eq6_13}
\end{equation}
where $\kappa>0$.
If condition (\ref{eq6_13}) is fulfilled, then
the constitutive equation for the diffusive flux ${\bf J}_d({\bf x},t)$
reduces to the Fickian expression ${\bf J}_d({\bf x},t)= - D_{\rm eff}
\, \nabla p({\bf x},t)$, where the effective diffusivity attains
the expression
\begin{equation}
D_{\rm eff}= \frac{2 \, \kappa \, D_{\rm nom}}{1-\delta}
\label{eq6_14}
\end{equation}

Next, consider the more general class of transitionally symmetric
GPK processes, where
the transition rates $\lambda_\alpha>0$ can be arbitrary,
and the matrix $K_{\alpha,\gamma}=
\lambda_\gamma \, A_{\alpha,\gamma}$  is symmetric.
No assumptions are made on the velocity vectors ${\bf b}_\alpha$.
Multiplying eq. (\ref{eq5_13}) by ${\bf b}_\alpha/\lambda_\alpha$
and summing over the states $\alpha$ provides
\begin{eqnarray}
\hspace{-2.3cm}
\partial_t \left [ \sum_{\alpha=1}^N \frac{ {\bf b}_\alpha \, p_\alpha({\bf x},t)}{\lambda_\alpha} \right ] & = & - \nabla \cdot \left [ {\bf v}({\bf x}) \, \sum_{\alpha=1}^N \frac{{\bf b}_\alpha \, p_\alpha({\bf x},t)}{\lambda_\alpha} \right ] -
\nabla \cdot \left [ \sum_{\alpha=1}^N \frac{ {\bf b}_\alpha \, {\bf b}_\alpha
\, p_\alpha({\bf x},t)}{\lambda_\alpha} \right ]
\nonumber \\
& = & - {\bf J}_d({\bf x},t) + \sum_{\alpha,\gamma=1}^N \frac{\lambda_\gamma}{\lambda_\alpha} \, {\bf b}_\alpha \, A_{\alpha,\gamma} \, p_\gamma({\bf x},t)
\label{eq6_15}
\end{eqnarray}
Assume that the stochastic velocity vectors ${\bf b}_\alpha$ satisfy
the transition conditions
\begin{equation}
\sum_ {\alpha=1}^N A_{\alpha,\gamma} \, \frac{{\bf b}_\alpha}{\lambda_\alpha}
= \delta \, \frac{{\bf b}_\gamma}{\lambda_\gamma} \, , \qquad
\gamma=1,\dots,N
\label{eq6_16}
\end{equation}
 which generalize
eq. (\ref{eq6_5}) to nonuniform $\lambda_\alpha$.
Eqs.  (\ref{eq6_16})
 represent
a system of $N$ constraints that complement  the $n$ zero-bias conditions
eq. (\ref{eq5_a5}), forming a system of $N+n$  linear  constraints to
be imposed on ${\bf b}_\alpha$ and $\lambda_\alpha$, $\alpha=1,\dots,N$.
Enforcing eq. (\ref{eq6_16}), it follows that:
\[
 \sum_{\alpha,\gamma=1}^N \frac{\lambda_\gamma}{\lambda_\alpha} \, {\bf b}_\alpha \, A_{\alpha,\gamma} \, p_\gamma({\bf x},t)
= \delta \, \sum_{\gamma=1}^N {\bf b}_\gamma \, p_\gamma({\bf x},t) = \delta \, {\bf J}_d({\bf x},t)
\]
Assuming that  $\delta \neq 1$,
 eq. (\ref{eq6_15}) becomes
\begin{eqnarray}
\hspace{-1cm}
(1-\delta) \, {\bf J}_d({\bf x},t) & = & - \left \{ \partial_t \left [
\sum_{\alpha=1}^N \frac{{\bf b}_\alpha \, p_\alpha({\bf x},t)}{\lambda_\alpha}
\right ] + \nabla \cdot \left [
{\bf v}({\bf x}) \, \sum_{\alpha=1}^N \frac{{\bf b}_\alpha \, p_\alpha({\bf x},t)}{\lambda_\alpha}
\right ] \right  \} \nonumber \\
& = & - \nabla \cdot \left [ \sum_{\alpha=1}^N \frac{{\bf b}_\alpha \,
{\bf b}_\alpha \, p_\alpha({\bf x},t)}{\lambda_\alpha} \right ]
\label{eq6_17}
\end{eqnarray}
Let us define $\widehat{{\bf b}}_\alpha={\bf b}_\alpha/ b^{(c)}$,
$\nu_\alpha = \lambda_\alpha /\lambda^{(c)}$, where
$b^{(c)}$ and $\lambda^{(c)}$ are given by eq. (\ref{eq5_9}), so that
eq. (\ref{eq6_17}) becomes
\begin{eqnarray}
\hspace{-2.0cm}
(1-\delta) \, {\bf J}_d({\bf x},t) & = & 
- \frac{b^{(c)}}{\lambda^{(c)}} \, 
\left \{ \partial_t \left [
\sum_{\alpha=1}^N \frac{\widehat{\bf b}_\alpha \, p_\alpha({\bf x},t)}{\nu_\alpha}
\right ] + \nabla \cdot \left [
{\bf v}({\bf x}) \, \sum_{\alpha=1}^N \frac{\widehat{\bf b}_\alpha \, p_\alpha({\bf x},t)}{\nu_\alpha}
\right ] \right  \} \nonumber \\
& = & -  \frac{(b^{(c)})^2}{\lambda^{(c)}} \,\nabla \cdot \left [ \sum_{\alpha=1}^N \frac{\widehat{\bf b}_\alpha \,
\widehat{\bf b}_\alpha \, p_\alpha({\bf x},t)}{\nu_\alpha} \right ]
\label{eq6_18}
\end{eqnarray}
In the Kac limit, corresponding to an  
infinitely fast recombination amongst the
partial probability  waves,  the local equilibrium
condition analogous to eq. (\ref{eq6_11}) can be assumed, namely
\begin{equation}
p_\alpha({\bf x},t)= \frac{p({\bf x},t)}{N} +
o(1/\lambda^{(c)}) \, , \qquad \alpha=1,\dots,N 
\label{eq6_19}
\end{equation}
Substituting this  asymptotic expression into
eq. (\ref{eq6_18}), one obtains, for $b^{(c)}, \lambda^{(c)} \rightarrow
\infty$, keeping constant  the nominal diffusivity, the expression
for the flux constitutive equation
\begin{equation}
{\bf J}_d({\bf x},t) = - \frac{2 \,  D_{\rm nom}}{1-\delta}
\, \nabla  \cdot \left [ \frac{1}{N} \sum_{\alpha=1}^N \frac{\widehat{\bf b}_\alpha \, \widehat{\bf b}_\alpha}{\nu_\alpha} \, p({\bf x},t) \right ] +
{\mathcal O}(1/\sqrt{\lambda^{(c)}})
\label{eq6_20}
\end{equation}
In order to achieve a consistent Kac limit, converging towards an isotropic
diffusion process, the structure  tensor $S_{h,k}$
\begin{equation}
S_{h,k}= \frac{1}{N} \sum_{\alpha=1}^N
\frac{\widehat{b}_{\alpha,h} \, \widehat{b}_{\alpha,k}}{\nu_\alpha}
= \kappa \, \delta_{h,k} \, \qquad h,k=1,\dots,n
\label{eq6_21}
\end{equation}
should be  isotropic
with $\kappa>0$.
If eq. (\ref{eq6_21}) is fulfilled, the effective diffusivity
of this general GPK scheme is still given by eq. (\ref{eq6_14})
with the difference that the structure factor $\kappa$ is now
defined via eq. (\ref{eq6_21}).

This completes the analysis of the Kac limit for GPK processes.
To conclude, it is worth mentioning a further generalization.
We have considered the Kac limit as an asymptotic convergence of
GPK processes towards an isotropic diffusion model. This
principle can be generalized, if needed, 
to encompass the convergence towards anisotropic 
Fickian diffusion. In the latter case, for $b^{(c)} \, , \lambda^{(c)}
\rightarrow \infty$, the Kac limit becomes a constitutive
equation of the form ${\bf J}_d({\bf x},t)= - {\mathbb D}_{\rm eff} \, \nabla p({\bf x},t)$, where ${\mathbb D}_{\rm eff}$ is a symmetric positive definite
tensor of effective diffusivities. The implications of this more 
general condition in terms of ${\mathcal B}_N$,
${\boldsymbol \Lambda}$ and ${\bf A}$ can  be developed
following the same approach applied in the fully isotropic case.

\subsection{Examples in ${\mathbb R}^2$}
\label{sec_5_2}

\begin{figure}[h]
\begin{center}
{\includegraphics[height=5.5cm]{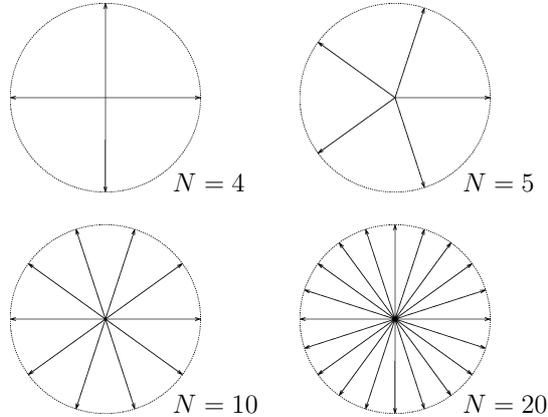}}
\end{center}
\caption{Examples of stochastic velocity
vectors defined by eq. (\ref{eq6_2_1}) for different values
of $N$.}
\label{Fig7}
\end{figure}

In this paragraph we apply the theory developed for Kac homogenization
using several examples in ${\mathbb R}^2$.
Consider the system of velocity vectors
\begin{equation}
{\bf b}_\alpha = b \,
\left  (
\begin{array}{c}
\cos(2 \pi (\alpha-1)/N)  \\
\sin(2 \pi (\alpha -1)/N)
\end{array}
\right )
\, , \;\;\; \alpha=1,\dots,N
\label{eq6_2_1}
\end{equation}
For $N$ even, this system is velocity symmetric.
Figure \ref{Fig7} shows  several examples of velocity
vectors obtained for different values of $N$.
This system of stochastic velocities satisfies the unbiasing
condition (\ref{eq5_a5}).
Let $\lambda_\alpha=\lambda$, uniformly with respect to
$\alpha=1,\dots,N$.

\begin{figure}[h]
\begin{center}
{\includegraphics[height=5.5cm]{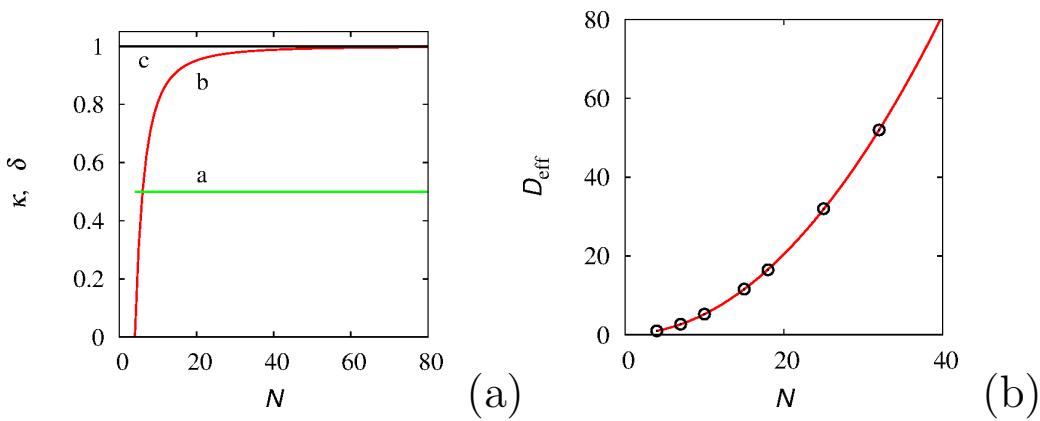}}
\end{center}
\caption{Panel (a): parameters $\kappa$ and $\delta$  vs $N$ for
the GPK model associated with eqs. (\ref{eq6_2_1})-(\ref{eq6_2_2})
with uniform $\lambda_\alpha$'s. Line (a) refers to
$\kappa=1/2$, line (b) to $\delta$, while line (c) depicts
the asymptotic limit $\delta=1$ for $N \rightarrow \infty$.
Panel (b): effective diffusivity $D_{\rm eff}$ vs $N$ for
$D_{\rm nom}=1$. Solid line refers to the theoretical prediction
eq.  (\ref{eq6_14}), symbols ($\circ$)  to the
results of stochastic simulations.}
\label{Fig8}
\end{figure}

As a first example let $A_{\gamma,\alpha}$  be given by
\begin{equation}
A_{\gamma,\alpha}= \frac{1}{2} \left ( \delta_{\gamma,[\alpha-1]}
+  \delta_{\gamma,[\alpha+1]} \right ) \, ,
\qquad \alpha,\gamma=1,\dots,N
\label{eq6_2_2}
\end{equation}
where $[\alpha]=\alpha$ for $\alpha=1,\dots,N$, $[0]=N$ and $[N+1]=1$.
Essentially, from state $\alpha$ transitions occurs towards the
two ``nearest neighboring'' states $[\alpha \pm 1]$ with equal probabilities.
This model satisfy the velocity transition condition (\ref{eq6_5})
for any $N\geq 2$. Moreover, the resulting structure
tensor is isotropic for $N \geq 3$.
Figure \ref{Fig8} panel (a) depicts the values
of the two structure parameters $\delta$ and $\kappa$ characterizing
this model as a function of the number $N$ of stochastic
velocity vectors considered. For this GPK process, $\kappa=1/2$
for any $N\geq 3$, while $\delta$ approaches $1$ as $N$ increases,
$1-\delta \sim N^{-2}$. Panel (b) of the same figure
depicts the predicted value of the effective diffusivity
given by eq. (\ref{eq6_14}) taking $D_{\rm nom}=1$. Symbols
($\circ$) corresponds to the estimate of $D_{\rm eff}$
from the scaling of the mean square displacements $\sigma_x^2(t)$,
$\sigma_y(t)$ along the $x$ and $y$ coordinates, ${\bf x}=(x,y)$,
$\sigma_x^2(t) \sim \sigma_y^2(t) \sim 2 \, D_{\rm eff} \, t$,
obtained from  stochastic simulations of eq. (\ref{eq5_4})
with ${\bf v}({\bf x})=0$, using an ensemble of $N=10^5$ particles.

As another example consider the symmetric matrix
\begin{equation}
A_{\gamma,\alpha}= \left \{
\begin{array}{ccc}
1/(N-1) & & \gamma \neq \alpha  \\
0 & & \gamma=\alpha
\end{array}
\right .
\, , \qquad  \alpha,\gamma=1,\dots,N
\label{eq6_2_3}
\end{equation}
keeping the same stochastic velocity vectors (\ref{eq6_2_1}).
This model corresponds to the model analyzed by Kolesnik \cite{kolesnik3}.
Since the ${\mathcal B}_N$ structure is unchanged, $\kappa=1/2$
for this model too.
Also in this case, the velocity transition conditions
(\ref{eq6_5}) are satisfied, with $\delta=-1/(N-1)$,
 and the  effective diffusivity,
using eq. (\ref{eq6_14}), is given by 
\begin{equation}
D_{\rm eff} = D_{\rm nom} \, \frac{N-1}{N}
\label{eq6_2_4}
\end{equation}
in agreement with the analysis by Kolesnik \cite{kolesnik3}.
If one slightly changes the transition probability matrix from
eq. (\ref{eq6_2_3}) to the
form
\begin{equation}
A_{\gamma,\alpha}= \frac{1}{N} \, , \qquad \alpha,\gamma=1,\dots,N
\label{eq6_2_5}
\end{equation}
one obtains $\delta=0$, providing $D_{\rm eff}=D_{\rm nom}$
for any $N \geq 3$.

\begin{figure}[h]
\begin{center}
{\includegraphics[height=5.5cm]{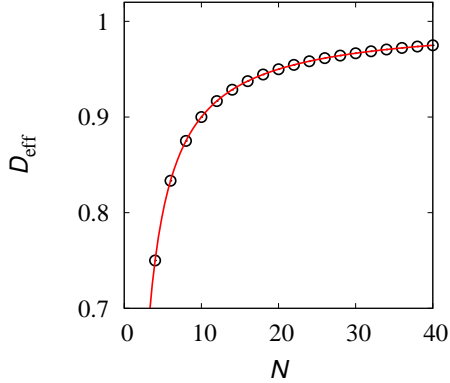}}
\end{center}
\caption{Effective diffusivity $D_{\rm eff}$ vs $N$
for the GPK model defined by eqs. (\ref{eq6_2_1}), (\ref{eq6_2_3}).
Solid line refers to eq. (\ref{eq6_2_4}), symbols
($\circ$) to the results of stochastic simulations. $D_{\rm nom}=1$.}
\label{Fig9}
\end{figure}
As can be observed, the fulfillment of the velocity transition condition
is a fairly common property of many GPK processes.

Next, consider  a system of nonuniform transition rates
$\lambda_\alpha$, $\alpha=1,\dots,N$, and a given transition
matrix $A_{\alpha,\gamma}$, and suppose that 
$\{ {\bf B}_\alpha \}_{\alpha=1}^N$ is a system of  unit vectors satisfying
the transition condition
\begin{equation}
\sum_{\alpha=1}^N A_{\alpha, \gamma} {\bf B}_\alpha =
\delta \, {\bf B}_\gamma \,  \qquad \gamma=1,\dots,N
\label{eq6_2_6}
\end{equation}
with $\delta \neq 1$.
Let
\begin{equation}
{\bf b}_\alpha = b \, \lambda_\alpha \, {\bf B}_\alpha
\, , \;\;\; \alpha=1,\dots N
\label{eq6_2_7}
\end{equation}
where $b>0$ is a parameter.
By definition, the vectors ${\bf b}_\alpha$ defined
by eq. (\ref{eq6_2_7}) satisfy the transition condition
eq. (\ref{eq6_16}), so that the further requirement to be imposed
on the transition rates $\lambda_\alpha$ is that  also the
zero-bias condition holds true, namely that
\begin{equation}
\sum_{\alpha=1}^N \lambda_\alpha \, {\bf B}_\alpha = 0
\label{eq6_2_8}
\end{equation}
Eq. (\ref{eq6_2_8}) represents a system of $n$ linear
constraints on the admissible values of the transition
rates $\lambda_\alpha$. Once a system of admissible
transition rates is determined via eq.  (\ref{eq6_2_8}),
the resulting GPK process admits an effective diffusivity tensor
${\mathbb D}_{\rm eff}$ the entries of which are given by
\begin{equation}
D_{{\rm eff},h,k}= \frac{1}{(1-\delta) \, N}
\, \sum_{\alpha=1}^N \frac{b_{\alpha,h} \, b_{\alpha,k}}{\lambda_\alpha}
= \frac{ 2 \, D_{\rm nom}}{(1-\delta ) \, N}
\sum_{\alpha=1}^N \nu_\alpha \, B_{\alpha,h} \, B_{\alpha,k} 
\label{eq6_2_9}
\end{equation}
where, in the present case, the nominal diffusivity
is $D_{\rm nom}= b^2 \lambda^{(c)}/2$. This
implies that the parameter $b$ should be chosen such that
$b \sim 1/\sqrt{\lambda^{(c)}}$.
In this way, anisotropic diffusion processes can be generated
in the Kac limit, corresponding to $\lambda^{(c)} \rightarrow \infty$.

\subsection{GPK in ${\mathbb R}^3$: construction of
${\mathcal B}_N$  from platonic solid geometry}
\label{sec_5_3}

In ${\mathbb R}^3$, the construction of suitable vector
systems ${\mathcal B}_N$, possessing enough
symmetries to ensure the convergence towards isotropic
diffusion in the Kac limit, can be grounded on the geometry of platonic solids.
As well known, there are five platonic (regular) solids, possessing
$N=4,\, 6,\, 8, \, 12,\, 20$ faces respectively: from $N=4$ corresponding
to the tetrahedron, up to $N=20$ for the icosahedron.

Given a platonic solid possessing $N$ faces, the system
of stochastic velocity vectors ${\mathcal B}_N$
can be constructed  by considering the normal
unit vectors to the $N$ faces of the solid, multiplied by 
a characteristic velocity $b^{(c)}$.

Consider uniform transition rates, i.e. $\lambda_\alpha=\lambda$,
$\alpha=1,\dots,N$ and the transition probability
matrix expressed either by eq. (\ref{eq6_2_4}) or by
eq. (\ref{eq6_2_5}).

For both the choices of the transition probability matrix, 
the transition conditions (\ref{eq6_5}) amongst the vectors
${\bf b}_\alpha$ are satisfied, and the resulting
diffusivity tensor is isotropic, namely
$D_{\rm eff,h,k}= \kappa \, D_{\rm nom} \, \delta_{h,k}$,
where $D_{\rm nom}= \left ( b^{(c)} \right )^2/(2 \lambda)$.
\begin{figure}[h]
\begin{center}
{\includegraphics[height=8cm]{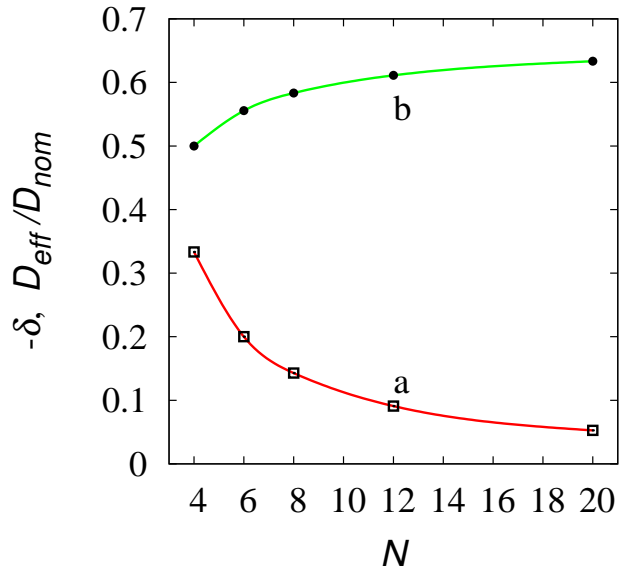}}
\end{center}
\caption{Behavior of the structural parameters $\delta$ and $\kappa=D_{\rm eff}/D_{\rm nom}$ for GPK processing defined from the geometry of the
regular platonic solids vs the number $N$ of faces, using the
transition probability matrix (\ref{eq6_2_3}).
Line (a) and ($\square$) refer to $-\delta$, line (b) and ($\bullet$)
to $D_{\rm eff}/D_{\rm nom}$.}
\label{Fig_plato}
\end{figure}

In the case of (\ref{eq6_2_5}), for all the five platonic
solids $\delta=0$ in eq. (\ref{eq6_5}), and $\kappa=1/3$.
In  the case of  eq. (\ref{eq6_2_3}) both $\delta$ and $\kappa$
depend on $N$, as depicted in figure \ref{Fig_plato}.
Observe that $\kappa$ is a monotonic function of $N$ attaining
the value $\kappa=1/2$ for $N=4$ (tetrahedron), up to
$\kappa\simeq 0.611$ for $N=20$ (icosahedron).

The generalization to higher dimensional spaces $n>3$ can be obtained
by applying the same procedure to higher dimensional polytopes.

\section{Homogenization: the Kac limit as 
emerging property}
\label{sec_6}

So far we have analyzed the Kac limit as a limit behavior
for $b^{(c)}$ and $\lambda^{(c)}$ tending to
infinity, keeping fixed the group $D_{\rm nom}$.
In point of fact, as purely stochastic dynamics is concerned,
i.e.,  for ${\bf v}({\bf x})=0$, the convergence
of GPK processes towards Brownian motion (Wiener processes)
 can be viewed as a long-term/large-distance
property occurring for any value of $b^{(c)}$ and $\lambda^{(c)}$.
The emerging Brownian-motion behavior characteristic of GPK processes is
analyzed in this Section, using moment analysis, which
is a classical technique in homogenization theory \cite{brenner}.
We show that in the long-term limit the statistical
characterization of a  GPK
process approaches the solution of a parabolic
model $\partial_t p= - {\bf V}_{\rm eff} \cdot \nabla p
+ \nabla \cdot \nabla \cdot  \left ( {\mathbb D}_{\rm eff} \, p \right )$,
where ${\bf V}_{\rm eff}$ and ${\mathbb D}_{\rm eff}$ are the
constant effective velocity vector and diffusivity tensor, respectively,
 determining
their values using the first elements of the moment hierarchy.
Strictly speaking we do not prove the convergence, but rather we
estimate ${\bf V}_{\rm eff}$ and ${\mathbb D}_{\rm eff}$
via first and second-order moment scaling. However,
the complete proof of the convergence towards a parabolic
model can be obtained using moment analysis and Pawula
theorem \cite{pawula}, estimating the property of the fourth-order
moments. This calculation is however extremely lengthy and
is left to the tenacious reader.

\subsection{Homogenization in the absence of a deterministic drift}
\label{sec_6_1}

Consider a GPK process in  the absence of a deterministic
drift, i.e., ${\bf v}(x)=0$.
In the  analysis  of the long-term properties
assume also  the following conditions:
(i) the matrix $K_{\alpha,\gamma}$ is symmetric;
(ii) the stochastic velocity vectors satisfy the zero-bias
condition (\ref{eq5_a5}).

The concept of long-term property can be made quantitative:
Consider the
 symmetric transfer matrix ${\bf A} \,\widehat{\boldsymbol \Lambda} -\widehat{\boldsymbol \Lambda}$ controlling the dynamics of state recombination.
 It admits $\mu^{(1)}=0$ as the dominant
eigenvalue, which is non degenerate (single multiplicity)
because the assumption of irreducibility. 
Let $\mu^{(2)}$ be the  second eigenvalue
with reversed sign. The characteristic time for relaxation
of the state dynamics of the $N$-state finite Poisson process
generating the dynamics is therefore $t_{\rm char}=(\mu^{(2)})^{-1}$,
so that long-term dynamics implies $t \gg t_{\rm char}$.

As stated above, consider a GPK process in the absence of a
deterministic drift, and the system
of lower-order partial moments
\begin{eqnarray}
m_\alpha^{(0)}(t) & = & \int_{{\mathbb R}^n} p_\alpha({\bf x},t)
\, d {\bf x} \nonumber \\
m_{\alpha,h}^{(1)}(t) & = & \int_{{\mathbb R}^n} x_h \, p_\alpha({\bf x},t)
\, d {\bf x}
\label{eq6_3_1} \\
m_{\alpha,h,k}^{(2)}(t) & = & \int_{{\mathbb R}^n} x_h \, x_k \, 
p_\alpha({\bf x},t) \, d {\bf x}
\nonumber
\end{eqnarray}
for $\alpha=1,\dots,N$, $h,k=1,\dots,n$.
The dynamics of the zero-th order partial moments is
given by
\begin{equation}
\partial_t m_\alpha^{(0)}(t) = - \lambda_\alpha \, m_\alpha^{(0)}(t)
+ \sum_{\gamma=1}^N \lambda_\gamma \, A_{\alpha,\gamma} \, m_\gamma^{(0)}(t)
\label{eq6_3_2}
\end{equation}
For $t \gg t_{\rm char}$, $m_\alpha^{(0)}(t)$, $\alpha=1,\dots,N$
 approach a
uniform distribution with respect to $\alpha$
\begin{equation}
m_\alpha^{(0)}(t)= \frac{1}{N} +{\mathcal O} \left (e^{-\mu^{(2)} \, t}
\right )
\label{eq6_3_3}
\end{equation}
which follows from the symmetry of ${\bf K}$.
The first-order moments are solution of the equations
\begin{equation}
\partial_t m_{\alpha,h}^{(1)}(t)= b_{\alpha,h} \, m_\alpha^{(0)}(t)
- \lambda_\alpha \, m_{\alpha,h}^{(1)}(t)
+ \sum_{\gamma=1}^N \lambda_\gamma \, A_{\alpha,\gamma} \, m_{\gamma,h}^{(1)}(t)
\label{eq6_3_3bis}
\end{equation}
Since $m_\alpha^{(0)}(t)$ converge to  constant values, the long-term limit
of the first-order moments is at most linear with time,
\begin{equation}
m_{\alpha,h}^{(1)}(t) = C_{\alpha,h}^{(1)} \, t + D_{\alpha,h}^{(1)}
+{\mathcal O} \left (e^{-\mu^{(2)} \, t}
\right )
\label{eq6_3_4}
\end{equation}
where $C_{\alpha,h}^{(1)}$ and $D_{\alpha,h}^{(1)}$,
$\alpha=1,\dots,N$, $h=1,\dots,n$ are
constant coefficients to be determined. Substituting eqs.
(\ref{eq6_3_4}) into eqs. (\ref{eq6_3_3bis}) the
following conditions amongst the expansion coefficients
are obtained in the long-time limit
\begin{equation}
-\lambda_\alpha \, C_{\alpha,h}^{(1)}
+ \sum_{\gamma=1}^N \lambda_\gamma \, A_{\alpha,\gamma} \, C_{\gamma,h}^{(1)}
=0
\label{eq6_3_5}
\end{equation}
\begin{equation}
C_{\alpha,h}^{(1)} = \frac{b_{\alpha,h}}{N} -\lambda_\alpha \, D_{\alpha,h}^{(1)}
+ \sum_{\gamma=1}^N \lambda_\gamma \, A_{\alpha,\gamma} \, D_{\gamma,h}^{(1)}
\label{eq6_3_6}
\end{equation}
Summing over $\alpha$ in eq. (\ref{eq6_3_6}), and
enforcing the zero-bias condition for the stochastic
velocity vectors one gets
\begin{equation}
\sum_{\alpha=1}^N C_{\alpha,h}^{(1)} = \frac{1}{N} \, \sum_{\alpha=1}^N
b_{\alpha,h} =0
\label{eq6_3_7}
\end{equation}
But the sum at the left-hand side of eq. (\ref{eq6_3_6})
is exactly the definition of the $h$-th entry of the
effective velocity vector, so that eq. (\ref{eq6_3_7}) implies
\begin{equation}
{\bf V}_{\rm eff} =0
\label{eq6_3_8}
\end{equation}
Moreover, by the symmetry  of the matrix $K_{\alpha,\gamma}=
\lambda_\gamma \, A_{\alpha,\gamma}$, it follows that
eq. (\ref{eq6_3_5}) is satisfied by a uniform
vector in $\alpha$, $C_{\alpha,h}^{(1)}=c_h$, and due to eq. (\ref{eq6_3_7}),
it follows that $c_h=0$. Therefore,
\begin{equation}
C_{\alpha,h}^{(1)}=0
\label{eq6_3_8bis}
\end{equation}
Next, consider the second-order partial moments $m_{\alpha,h,k}^{(2)}(t)$
that satisfy the system of evolution equations
\begin{equation}
\hspace{-1.5cm}
\partial_t m_{\alpha,h,k}^{(2)}(t)= b_{\alpha,h} \, m_{\alpha,k}^{(1)}(t)
+ b_{\alpha,k} \, m_{\alpha,h}^{(1)}(t)
- \lambda_\alpha \, m_{\alpha,h,k}^{(2)}(t)
+ \sum_{\gamma=1}^N \lambda_\gamma \, A_{\alpha,\gamma} \, m_{\gamma,h,k}^{(2)}(t)
\label{eq6_3_9}
\end{equation}
Due to the long-term properties of $m_{\alpha,h}^{(1)}(t)$, 
the second-order moments are, for $t \gg t_{\rm char}$,
at most quadratic in time, to the leading order
\begin{equation}
m_{\alpha,h,k}^{(2)}(t) = E_{\alpha,h,k}^{(2)} \, t^2 + F_{\alpha,h,k}^{(2)}
\, t+ G_{\alpha,h,k}^{(2)} +{\mathcal O} \left (e^{-\mu^{(2)} \, t}
\right )
\label{eq6_3_10}
\end{equation}
In point of fact, since $C^{(1)}_{\alpha,h}=0$, it follows that
the coefficients $E_{\alpha,h,k}^{(2)}$ are identically vanishing.
At present, we do not need to enforce this property.
The substitution of this expression into
the balance equation (\ref{eq6_3_9}) provides the
system of relations amongst the expansion coefficients
\begin{equation}
-\lambda_\alpha \, E_{\alpha,h,k}^{(2)}
+ \sum_{\gamma=1}^N \lambda_\gamma \, A_{\alpha,\gamma} \, E_{\gamma,h,k}^{(2)}
=0
\label{eq6_3_11}
\end{equation}
\begin{equation}
2 \,E_{\alpha,h,k}^{(2)} = b_{\alpha,h}
\, C_{\alpha,k}^{(1)} + b_{\alpha,k}
\, C_{\alpha,hs}^{(1)} -\lambda_\alpha \, F_{\alpha,h,k}^{(2)}
+ \sum_{\gamma=1}^N \lambda_\gamma \, A_{\alpha,\gamma} \, F_{\gamma,h,k}^{(2)}
\label{eq6_3_12}
\end{equation}
\begin{equation}
F^{(2)}_{\alpha,h,k} = b_{\alpha,h}
\, D_{\alpha,k}^{(1)} + b_{\alpha,k}
\, D_{\alpha,h}^{(1)} -\lambda_\alpha \, G_{\alpha,h,k}^{(2)}
+ \sum_{\gamma=1}^N \lambda_\gamma \, A_{\alpha,\gamma} \, G_{\gamma,h,k}^{(2)}
\label{eq6_3_13}
\end{equation}

At this stage it is necessary to assume some
conditions on the stochastic velocity vectors ${\bf b}_\alpha$.
As done in Section \ref{sec_5}, let us assume the velocity transition
condition  (\ref{eq6_16}) with $\delta \neq 1$.

Summing over $\alpha$ in eq. (\ref{eq6_3_11}) provides
\begin{equation}
2 \sum_{\alpha=1}^N E^{(2)}_{\alpha,h,k}=
\sum_{\alpha=1}^N b_{\alpha,k} C^{(1)}_{\alpha,h}
+ \sum_{\alpha=1}^N b_{\alpha,h} C^{(1)}_{\alpha,k} =0
\label{eq6_3_14}
\end{equation}
The right-hand side in eq. (\ref{eq6_3_14}) is
identically vanishing, because of eq. (\ref{eq6_3_8bis}).
But the term at the left-hand side of eq. (\ref{eq6_3_14}) is
just the prefactor of the quadratic contribution associated
with the overall second-order moment $m_{h,k}^{(2)}=\sum_{\alpha=1}^N
m_{\alpha,h,k}^{(2)}$.
Since the first-order moments are asymptotically constant,
eq. (\ref{eq6_3_14}) implies that the overall covariance
matrix $\sigma_{h,k}^2(t)= \sum_{\alpha=1}^N m_{\alpha,h,k}^{(2)}(t)-
 \sum_{\alpha=1}^N m_{\alpha,h}^{(1)}(t) \sum_{\gamma=1}^N m_{\gamma,k}^{(1)}(t)$  scales in the long-term regime linearly with time $t$, i.e.,
\begin{equation}
\sigma_{h,k}^2(t) \sim \sum_{\alpha=1}^N m_{\alpha,h,k}^{(2)}(t)
\sim \left ( \sum_{\alpha=1}^N F^{(2)}_{\alpha,h,k} \right ) t
= 2 D_{{\rm eff},h,k} \, t
\label{eq6_3_15}
\end{equation}
and that the sum over the states of $F_{\alpha,h,k}^{(2)}$ is
twice the entry $D_{{\rm eff},h,k}$ of the effective diffusivity
tensor ${\mathbb D}_{\rm eff}$.
To obtain an expression for this quantity, use
eq. (\ref{eq6_3_13})
\begin{equation}
\sum_{\alpha=1}^N F_{\alpha,h,k}^{(2)} =
\sum_{\alpha=1}^N b_{\alpha,h} \, D_{\alpha,k}^{(1)} +
\sum_{\alpha=1}^N b_{\alpha,k} \, D_{\alpha,h}^{(1)}
\label{eq6_3_16}
\end{equation}
The term at the right-hand side can be estimated
enforcing eqs. (\ref{eq6_3_6}) and (\ref{eq6_3_8bis}),
derived from the dynamics of the first-order moments.
Multiplying eq. (\ref{eq6_3_6}) by $b_{\alpha,k}/\lambda_\alpha$,
summing over the states $\alpha$, and enforcing the
velocity transition condition (\ref{eq6_16}) one gets
\begin{eqnarray}
0 & = & \frac{1}{N} \sum_{\alpha=1}^N \frac{b_{\alpha,h} \, b_{\alpha,k}}{\lambda_\alpha} - \sum_{\alpha=1}^N b_{\alpha,k} \, D_{\alpha,h}^{(1)}
+ \sum_{\alpha,\gamma=1} \frac{\lambda_\gamma \, b_{\alpha,k}}{\lambda_\alpha}
\, A_{\alpha,\gamma} \, D_{\gamma,h}^{(1)} \nonumber \\
& = & \frac{1}{N} \sum_{\alpha=1}^N \frac{b_{\alpha,h} \, b_{\alpha,k}}{\lambda_\alpha} 
- (1-\delta) \sum_{\alpha=1}^N b_{\alpha,k} \, D_{\alpha,h}^{(1)}
\label{eq6_3_17}
\end{eqnarray}
which provides for $\sum_{\alpha=1}^N F_{\alpha,h,k}^{(2)}$ 
the expression
\begin{equation}
\sum_{\alpha=1}^N F_{\alpha,h,k}^{(2)}
=  \frac{2}{N \, (1-\delta)} \sum_{\alpha=1}^N \frac{b_{\alpha,h} \, b_{\alpha,k}}{\lambda_\alpha} 
\label{eq6_3_18}
\end{equation}
and for $D_{{\rm eff},h,k}$ an expression identical to
eq. (\ref{eq6_20}).

Some concluding observations on the long-term properties
follows from the moment analysis
addressed in this paragraph. In the case of purely stochastic
GPK processes, the asymptotic approach towards a Brownian-like
behavior is not solely a mathematical property that holds 
in the Kac limit, letting $b^{(c)}$ and $\lambda^{(c)}$ diverge,
keeping fixed $D_{\rm nom}$. It is  also a long-term asymptotics
of the process. This makes GPK processes particularly appealing
in the description of the physical systems in all the
cases where the stochastic fluctuations occur
at characteristic
time-scales definitely shorter than the observation time-scales.
It is also remarkable that all the conditions derived in Section \ref{sec_5}
have been naturally and simply re-derived using homogenization
techniques in the long-term limit.

\subsection{A generalization}
\label{sec_6_2}

Within the framework of homogenization theory it is
comfortable to address the  case where no velocity
transition conditions are assumed, and the stochastic velocity vectors
fulfills solely the zero-bias condition (\ref{eq5_a5}).
Even in this more general setting it is possible
to derive a long-term Brownian behavior of GPK processes.
Enforcing the fact that the first-order partial moments
saturates asymptotically, i.e., $C_{\alpha,h}^{(1)}=0$,
it follows that $E_{\alpha,h,k}^{(2)}=0$ identically so
that the effective diffusivity tensor can be defined and,
by eqs.  (\ref{eq6_3_13})-(\ref{eq6_3_14}), is 
expressed by
\begin{equation}
D_{{\rm eff},h,k}= \frac{1}{2} \sum_{\alpha=1}^N \left (
b_{\alpha,h} \, D_{\alpha,k}^{(1)} + b_{\alpha,k} \, D_{\alpha,h}^{(1)}
\right )
\label{eq6_3_19}
\end{equation}

 From eqs. (\ref{eq6_3_6}) and (\ref{eq6_3_8bis})
it follows that, for a fixed value of the index $h$, the coefficients $D_{\alpha,h}^{(1)}$
satisfy a linear system of equations
\begin{equation}
({\bf K}-\widehat{\boldsymbol \Lambda} ) \, {\bf D}^{(1)}_h = {\bf f}_h
\, , \qquad h=1,\dots,n
\label{eq6_3_20}
\end{equation}
where we have set ${\bf D}^{(1)}_h=(D_{1,h}^{(1)},\dots,D_{N,h}^{(1)})$
and ${\bf f}_h = (-b_{1,h},\dots,-b_{N,h})/N$.
Eqs. (\ref{eq6_3_20}) represent $n$ decoupled systems
of linear equations for ${\bf D}^{(1)}_h$, $h=1,\dots,n$,
the coefficient matrices of which are singular.

The matrix ${\bf K}-\widetilde{\boldsymbol{\Lambda}}$ is,
by assumption, symmetric, so that it possesses
a system of real eigenvalues $\{\mu^{(\beta)}\}_{\beta=1}^N$
ordered such that $\mu^{(1)}>\mu^{(2)} \geq \cdots$,
and of eigenfunctions $\{\boldsymbol{\psi}^{(\beta)}\}_{\beta=1}^N$,
\begin{equation}
\left ( {\bf K}-\widetilde{\boldsymbol{\Lambda}} \right ) \, \boldsymbol{\psi}^{(\beta)} = \mu^{(\beta)} \, \boldsymbol{\psi}^{(\beta)}
\,, \qquad \beta=1,\dots,N
\label{eq6_ad_1}
\end{equation}
that can be always assumed to be mutually orthogonal and normalized
to unit norm. The assumption of irreducibility implies
that $\mu^{(1)}=0$ is non degenerate, so that $\mu^{(\beta)}<0$,
for $\beta \geq 2$. Specifically, the dominant
eigenfunction is given by $\boldsymbol{\psi}^{(1)}=\frac{1}{\sqrt{N}}
(1,\dots,1)$.

 Let $({\bf f},\boldsymbol{\psi})$ the usual
Euclidean scalar product for $N$-dimensional vectors ${\bf f}$,
$\boldsymbol{\psi}$.  The zero-bias condition
on the velocity vectors ${\bf b}_\alpha$ implies that
\begin{equation}
({\bf f}_h , \boldsymbol{\psi}^{(1)}) =
\frac{1}{\sqrt{N}} \sum_{\alpha=1}^N f_{\alpha,h}
= - \frac{1}{N^{3/2}}  \sum_{\alpha=1}^N b_{\alpha,h} =0
\label{eq6_ad_2}
\end{equation}
for any $h=1,\dots, n$.
Therefore, the forcing terms  ${\bf f}_h$ in the
linear systems (\ref{eq6_3_20}) admit a vanishing projection onto
the subspace spanned by the dominant (and vanishing) eigenvalue,
and this ensures that eqs. (\ref{eq6_3_20}) admit a solution,
and indeed infinitely many.
The solution of eq. (\ref{eq6_3_20}) can be expressed
by respect to the eigenfunctions of the coefficient matrix
as
\begin{equation}
{\bf D}_h^{(1)} = c_h^{(1)} \, \boldsymbol{\psi}^{(1)}
+ \sum_{\beta=2}^N c_h^{(\beta)} \, \boldsymbol{\psi}^{(\beta)}
\label{eq_6_ad_3}
\end{equation}
Substituting this expansion into eq. (\ref{eq6_3_20})
after elementary algebra follows that for $\beta\geq 2$
the expansion coefficients $c_h^{(\beta)}$ are
univocally defined and given by
\begin{equation}
c_h^{(\beta)}= \frac{({\bf f}_h, \boldsymbol{\psi}^{(\beta)})}{\mu^{(\beta)}}
\, , \qquad \beta=2,\dots,N
\label{eq6_ad_4}
\end{equation}
so that
\begin{equation}
D_{\alpha,h}^{(1)} = \frac{(c_h^{(1)}/\sqrt{N}
+ \sum_{\beta=2}^N  \boldsymbol{\psi}^{(\beta)})}{\mu^{(\beta)}}
\, \psi_\alpha^{(\beta)}
\label{eq6_ad_5}
\end{equation}
depending on the arbitrary constants $c_h^{(1)}$.
The first term at the r.h.s of eq. (\ref{eq6_ad_5})
provides a vanishing contribution to
the expression of the effective tensor diffusivity
(\ref{eq6_3_19}), due to the zero-bias conditions,
so that the substitution of eq. (\ref{eq6_ad_5})
into eq. (\ref{eq6_3_19}) provides the following
expression for $D_{{\rm eff},h,k}$
\begin{equation}
D_{{\rm eff},h,k}=
\frac{1}{2} \sum_{\alpha=1}^N \sum_{\beta=2}^N  \frac{\psi_\alpha^{(\beta)}}{\mu^{(\beta)}}
\, \left [ b_{\alpha,h} \, ({\bf f}_k,\boldsymbol{\psi}^{(\beta)})
+ b_{\alpha,k} \,  ({\bf f}_h,\boldsymbol{\psi}^{(\beta)}) \right ]
\label{eq6_ad_6}
\end{equation}
representing the formal solution for the effective diffusivity
tensor in the general case.
From the above derivation it can be observed that, while
the zero-bias condition represents a fundamental requisite
in order to achieve an emergent Brownian behavior, the assumption
of a given  velocity transition condition is simply a very
convenient property that provides an easy and compact expression
for the Kac limit and for the emerging long-term properties,
but that is not necessary for their occurrence, albeit
in this general case the effective tensor diffusivity
may not be isotropic.

\subsection{Presence of a deterministic drift}
\label{sec_6_3}

In the previous paragraphs
 we have not considered the presence of a deterministic drift
${\bf v}({\bf x})$.
 This is because its presence may change completely
the qualitative properties of GPK processes for small/moderate
values of $b^{(c)}$ and $\lambda^{(c)}$. 

The presence of a deterministic bias breaks down the symmetry
between the Kac-limit in the parameter space and the
long-term limit of GPK processes over all the range of
parameters $b^{(c)}$ and $\lambda^{(c)}$.
In point of fact this phenomenon can occur both for solenoidal
and irrotational deterministic biasing fields ${\bf v}({\bf x})$.

Particularly critical is the case where ${\bf v}({\bf x})$
is irrotational, i.e., it stems from a potential.
In this case, it may happen that for small values of
$b^{(c)}$ and $\lambda^{(c)}$ the dynamics of a GPK process is not even ergodic, as it becomes trapped into many different invariant domains,
each of which characterized by a unique stationary invariant
probability density function.
The phenomenon of ergodicity breaking is analyzed in \cite{giona_epl}
for one-dimensional Poisson-Kac dynamics, and thoroughly addressed
in part II for higher dimensional systems.

\subsection{A degenerate case: transitionally deterministic GPK processes}
\label{sec_6_4} 

The assessment of the Kac limit for GPK processes
goes well beyond the class of examples considered
in paragraph \ref{sec_6_2}. In this paragraph we
consider another example that is interesting for
highlighting other properties of GPK processes,
represented by GPK models that are not transitionally symmetric,
and that are associated with a rather peculiar transition
mechanism: transition from any state occurs just towards a single
one.  For this reason, these models can be referred
to as {\em transitionally deterministic}.
More precisely, a GPK process is transitionally
deterministic if there exists a
one-to-one mapping $\varphi: \{1,\dots,N\} \rightarrow \{1,\dots,N \}$
such that
\begin{equation}
A_{\beta,\alpha}= \delta_{\beta,\varphi(\alpha)} \, ,
\;\;\; \alpha,\beta=1,\dots,N
\label{eq6_4_1}
\end{equation}
A transitionally deterministic process
 is irreducible, if the iterates of $\varphi$ starting from
any state $\alpha$ spam the whole set of possible states
$\{1,\dots,N\}$. 

A peculiar case of transitionally deterministic GPK processes
is the one-dimensional  Poisson-Kac process analyzed in Section \ref{sec_2},
where the stochastic forcing $b\,  (-1)^{\chi(t)}$ is strictly
dichotomic. This occurs because the number of distinct states $N$
equals $2$. In this case, the process is also transitionally
symmetric. The latter property however does not hold
for $N >2$.

Transitionally deterministic processes are interesting for
the following reason: 
albeit the strictly deterministic transition mechanism, the
long-term dynamics possesses a dissipative character, corresponding
to the Kac limit. This is essentially due to the continuous
time parametrization, which implies that,  at any time $t$,
there is a finite
probability of remaining at a given state $\alpha$,  not  
switching to the state $\varphi(\alpha)$, and this  provides
an asymptotic dispersive behavior.

Below, we consider a subclass of these processes for which:
(i) all the transition rates are equal $\lambda_\alpha=\lambda$,
(ii) the velocity vectors ${\bf b}_\alpha$ satisfy the zero bias
condition  eq. (\ref{eq5_a5}), and finally, (iii) the process
is velocity symmetric with the property that there exists
an integer $N_i<N$ such that
\begin{equation}
{\bf b}_{\varphi^{N_i}(\alpha)} = - {\bf b}_\alpha \, , \;\;\;
\alpha=1,\dots,N
\label{eq6_4_2}
\end{equation}
where $\varphi^m(\alpha)$ is the $m$-th iteration 
of the map $\varphi$ starting from state $\alpha$.
Condition (iii) is not necessary to provide a Kac limit, but
simplifies significantly the analysis. 

The balance equations for the partial probability waves
$p_\alpha({\bf x},t)$ in transitionally deterministic
GPK processes, characterized by uniform transition rates,
read
\begin{equation}
\hspace{-2.0cm}
\partial_t p_\alpha({\bf x},t) = - \nabla \cdot \left [
 {\bf v}({\bf x}) \, p_\alpha({\bf x},t)
\right ] - \nabla \cdot \left [ {\bf b}_\alpha \, p_\alpha({\bf x},t) \right ]
- \lambda \, \left [p_\alpha({\bf x},t) - p_{\psi(\alpha)({\bf x},t)} \right ] 
\label{eq6_4_3}
\end{equation}
where $\psi=\varphi^{-1}$ is the inverse mapping of $\varphi$.
The diffusive flux ${\bf J}_d({\bf x},t)$ is the solution of the
equation
\begin{eqnarray}
\partial_t {\bf J}_d({\bf x},t) & = & - \nabla \cdot \left [ {\bf v}({\bf x})
 \, {\bf J}_d({\bf x},t) \right ]
- \nabla \cdot \left [ \sum_{\alpha=1}^N {\bf b}_\alpha \, {\bf b}_\alpha
\, p_\alpha({\bf x},t) \right ] 
\nonumber \\
& - & \lambda \, {\bf J}_d({\bf x},t)
 + \lambda \, {\bf J}_{d,1}({\bf x},t)
\label{eq6_4_4}
\end{eqnarray}
where
\begin{equation}
{\bf J}_{d,1}({\bf x},t)= \sum_{\alpha=1}^N {\bf b}_\alpha \,
p_{\psi(\alpha)}({\bf x},t)= \sum_{\alpha=1}^N {\bf b}_{\varphi(\alpha)}
\, p_\alpha({\bf x},t) 
\label{eq6_4_5}
\end{equation}
Set
\begin{equation}
{\bf J}_{d,m}({\bf x},t)=  \sum_{\alpha=1}^N {\bf b}_{\varphi^m(\alpha)}
\, p_\alpha({\bf x},t) \, , \;\;\; m=0,1,\dots
\label{eq6_4_6}
\end{equation}
where ${\bf J}_{d,0}({\bf x},t)={\bf J}_d({\bf x},t)$.
The balance equation for ${\bf J}_{d,m}({\bf x},t)$, $m>0$,
can be obtained multiplying eq. (\ref{eq6_4_3}) by
${\bf b}_{\varphi^m(\alpha)}$ and summing over $\alpha$
\begin{eqnarray}
\partial_t {\bf J}_{d,m}({\bf x},t)
& = & - \nabla \cdot \left [  {\bf v}({\bf x}) \, {\bf J}_{d,m}({\bf x},t) 
\right ] 
- \nabla \cdot \left [ \sum_{\alpha} {\bf b}_{\varphi^m(\alpha)}
 \, {\bf b}_\alpha
\, p_\alpha({\bf x},t) \right ] 
\nonumber \\
&- & \lambda \, {\bf J}_{d,m}({\bf x},t) + \lambda \,
 {\bf J}_{d,m+1}({\bf x},t)
\label{eq6_4_7}
\end{eqnarray}
Summing over $m$ from $m=0$ up to $m=M>0$ 
\begin{eqnarray}
\frac{1}{\lambda} \left \{ \sum_{m=0}^M \left [
\partial_t {\bf J}_{d,m}({\bf x},t) + \nabla \cdot \left (
  {\bf v}({\bf x}) \, {\bf J}_{d,m} ({\bf x},t)
\right ) \right ] \right \}  = 
\nonumber \\
-  2 \, D_{\rm nom} \nabla \cdot
\left [ \sum_{m=0}^M \sum_{\alpha=1}^N 
\widehat{\bf b}_{\varphi^m(\alpha)}
 \, \widehat{\bf b}_\alpha
\, p_\alpha({\bf x},t) \right ] 
- {\bf J}_d({\bf x},t) +{\bf J}_{d,M+1}({\bf x},t)
\label{eq6_4_8}
\end{eqnarray}
where we have set ${\bf b}_\alpha=b^{(c)} \, \widehat{\bf b}_{\alpha}$,
$D_{\rm nom}=\left (b^{(c)} \right )^2/2 \lambda$. 
Enforcing property (iii), it follows that 
${\bf J}_{d,N_i}({\bf x},t)={\bf J}_d({\bf x},t)$. Therefore, setting $M=N_i-1$ in eq. (\ref{eq6_4_8}),
and enforcing the  infinitely fast partial-wave recombination, $p_\alpha({\bf x},t)=p({\bf x},t)/N+o(\lambda^{-1})$,
one obtains the constitutive equation for the diffusive flux in the
Kac limit 
\begin{equation}
{\bf J}_d({\bf x},t)= - D_{\rm nom} \, \nabla \cdot \left [
{\bf S} \, p({\bf x},t) \right ]
\label{eq6_4_9}
\end{equation}
where, in the present case, the structure tensor ${\bf S}$ is given by
\begin{equation}
{\bf S}= \frac{1}{N} \,  \sum_{m=0}^{N_i-1} \sum_{\alpha=1}^N 
\widehat{\bf b}_{\varphi^m(\alpha)}
 \, \widehat{\bf b}_\alpha
\label{eq6_4_10}
\end{equation}
Observe that the structure tensor defined by
eq. (\ref{eq6_4_9}) needs not to be symmetric.
Eqs. (\ref{eq6_4_9})-(\ref{eq6_4_10}) provide the expression for the Kac limit
of this class of GPK processes.

\begin{figure}[h]
\begin{center}
{\includegraphics[height=8cm]{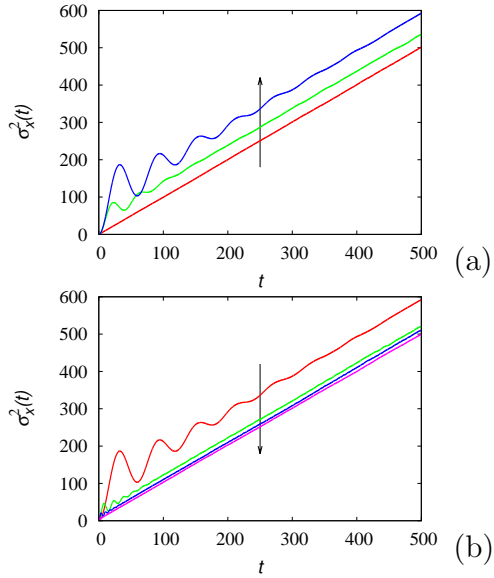}}
\end{center}
\caption{$\sigma_x^2(t)$ vs $t$ for the transitional
deterministic GPK process discussed in the main
text. Panel (a): $b=1$. The arrow indicates increasing values of $N=4,\, 20,
\, 30$. Panel (b): $N=30$. The arrows indicates increasing values
of $b=1,\,2,\,3,\,5$.} 
\label{Fig10}
\end{figure}

Consider a practical example in ${\mathbb R}^2$. Let
the stochastic velocity vectors be given by eq. (\ref{eq6_2_1})
with $b^{(c)}=b$,
and  the mapping $\varphi$
be 
\begin{equation}
\varphi(\alpha)=
\left \{
\begin{array}{ccc}
\alpha+1 & & \alpha=1,N-1 \\
1 & & \alpha=N
\end{array}
\right .
\label{eq6_4_11}
\end{equation}
For $N$ even and greater than 2, property (iii) is satisfied by
$N_i=N/2$. In this case, the direct calculation of the
entries of the structure tensor provides
\begin{equation}
S_{1,1}=S_{2,2}= \frac{1}{2} \, , \;\;\;
S_{2,1}=-S_{1,2}
\label{eq6_4_12}
\end{equation}
so that the resulting GPK process admits an isotropic Kac limit
with
\begin{equation}
D_{\rm eff}= \frac{D_{\rm nom}}{2}
\label{eq6_4_13}
\end{equation}
Figure \ref{Fig10} depicts the mean square displacement
$\sigma_x^2(t)$ vs $t$ obtained from stochastic simulations
of this GPK process in the absence of a deterministic bias, i.e.,
${\bf v}({\bf x})=0$ (an identical result can be obtained
if $\sigma_y^2(t)$ would be considered) at $D_{\rm nom}=1$. Panel (a)
addresses the case $b=1$ at different values of $N$,
while panel (b) depicts the Kac limit for $N=30$ at increasing
values of $b$, keeping fixed $D_{\rm nom}=1$.
The following observations follow from the inspection
of these data: (i) the estimate (\ref{eq6_4_13})
is in perfect agreement with the numerical simulations,
as for $D_{\rm nom}=1$, $D_{\rm eff}=1/2$
and thus $\sigma_x^2(t) \sim 2 D_{\rm eff} t = t$;
(ii)  also for this class of processes, the Brownian-motion limit 
corresponds to an emerging long-term property,
 occurring for any value of $b$ and $\lambda$,
although the characteristic time to reach this limit strongly depends on $N$,
and significantly increases with the number of states considered;
(iii) at short-time scales, and for high values
of $N$, $\sigma_x^2(t)$ exhibits
an oscillatory behavior, deriving from
the circulant nature of the transition matrix which possesses
a complex conjugate second eigenvalue. 

\section{Concluding remarks}
\label{sec_conclusions}

GPK processes represent a very versatile generalization
of the original Kac's model of telegrapher's noise extended
to higher dimensional systems. They
provide a broad class of stochastic processes 
possessing finite propagation velocity and, consequently,
smooth (almost everywhere differentiable) trajectories.
Their statistical description involves $N$ partial probability
density functions, satisfying a hyperbolic system of equations
with partial wave recombination, modulated by the transition
probability matrix $A_{\alpha,\beta}$ 
and by the transition rates $\lambda_\alpha$. 
This ``augmented'' statistical description
can be applied to ground, on stochastic microdynamic basis, 
the transport models of extended thermodynamic theory (see part III).

In this first part of the work we have analyzed the structural
properties of GPK processes, i.e., how the
structural features of the stochastic velocity vectors
${\mathcal B}_N = \{ {\bf b}_\alpha \}_{\alpha=1}^N$,
of the transition probability matrix ${\bf A}=(A_{\alpha,\beta})_{\alpha,\beta=1}^N$ and of the transition rate vector $\boldsymbol{\Lambda}=(\lambda_{\alpha})_{\alpha=1}^N$ impact on the Kac limit and on the the long-term
properties of the dynamics.

In the absence of deterministic biasing fields, there
is a strict correspondence between the Kac-limit behavior
occurring for $b^{(c)},\lambda^{(c)} \rightarrow \infty$,
keeping fixed the value of $D_{\rm nom}$, and the
long-term behavior for any finite values of $b^{(c)}$ and
$\lambda^{(c)}$, providing the same value of $D_{\rm nom}$,
for time scales $t \gg 1/\lambda^{(c)}$. Therefore,
the Kac limit (and specifically the convergence towards
classical parabolic transport schemes) is an emergent
feature of GPK models.

Attention has been focused on the Kac limit which
is a fundamental requisite connecting GPK processes
to classical stochastic models driven by Wiener perturbations.
A useful tool to derive the Kac limit behavior has been 
the transition condition (\ref{eq6_5}) connecting the
transition probability matrix to the structure
of the velocity vectors ${\bf b}_\alpha$.

All the physical implications of GPK dynamics and their
extensions/generalizations are thoroughly addressed in 
part II and III of this work.

\appendix
\section{Appendix}
\label{sec_appendix}

The cornerstone of the theory of hyperbolic and undulatory
transport is represented by the
model proposed by Kac \cite{kac} in one spatial dimension,
and referred by the author as ``telegrapher's noise''.
It is a Langevin equation where, instead of a stochastic
perturbation expressed as the increment of a Wiener
process, a Poisson-driven fluctuation dynamics is considered.

Let $\chi(t)$, $t \geq 0$, be a Poisson process, describing
the number of events of
a stationary, memoryless, ordinary stochastic process
in the time interval $[0,t)$.
The process is described by the probability $\pi(t;\Delta t)$
that an event occurs in the time interval $[t,t+\Delta t)$, given
by
\begin{equation}
\pi(t;\Delta t) = \lambda \, \Delta t + o(\Delta t)
\label{eq2_1}
\end{equation}
where $\lambda>0$. It follows from eq. (\ref{eq2_1})
the balance equations for the probabilities $P_n(t)=\mbox{Prob}\left [
\chi(t)=n \right ]$, $n=0,1,\dots$
\begin{eqnarray}
\dot{P}_0(t) & = & -\lambda \, P_0(t) \nonumber \\
\dot{P}_n(t) & = & -\lambda \, P_{n}(t) + \lambda \, P_{n-1}(t)
\label{eq2_2}
\end{eqnarray}
where $\dot{P_n}(t)=d P_n(t)/dt$,
out of which the Poissonian statistics, follows
setting $P_0(0)=1$.

Let $v(x)$ a deterministic velocity field $ x \in (-\infty,\infty)$,
$b>0$ a characteristic
velocity, and consider the stochastic differential equation
\begin{equation}
d x(t) = v(x(t)) \, dt + b \, (-1)^{\chi(t)} \, dt
\label{eq2_3}
\end{equation}
equipped with the initial condition $x(t_0=0)=x_0$, where $\chi(t)$
is the above mentioned Poisson process. 
The stochastic perturbation in eq. (\ref{eq2_3})
acts by switching $(-1)^{\chi(t)}$ from $+1$, to $-1$.
Eq. (\ref{eq2_3}) is referred to as the
Poisson-Kac equation, and the associated process as the
Poisson-Kac process.

Since the stochastic perturbation should not possess any bias for any $t\geq 0$, it is
natural to assume the following initial conditions for the
Poisson process
\begin{equation}
\mbox{Prob} \left [ \chi(0)=0 \right ] =\mbox{Prob}
\left [ \chi(0)=1 \right ]= \frac{1}{2}
\label{eq2_4}
\end{equation}
and zero otherwise, i.e., $P_n(0)=0$, $n>1$.
In such as way, $ E \left [ (-1)^{\chi(t)} \right ]=0$
for all $t\geq 0$, where $E[\cdot]$ indicates  expectation values.
From the initial condition (\ref{eq2_4}), it follows that
\begin{equation}
P_n(t)= \frac{1}{2} \frac{(\lambda \, t)^n \, e^{-\lambda \, t}}{n!}
+  \frac{1}{2} \frac{(\lambda \, t)^{n-1} \, e^{-\lambda \, t}}{(n-1)!}
\, \eta(n-1)
\label{eq2_5}
\end{equation}
where $\eta(n)$ is a discrete Heaviside function
$\eta(n)=1$ for $n \geq 0$, $\eta(n)=0$ for $n<0$.

Therefore, $E \left [ \chi(t) / \chi(0)=0 \right ] = \lambda \, t$,
$E\left [ \chi(t) / \chi(0)=1 \right ]=\lambda \, t + 1$, and
 the correlation function for $t, \tau \geq 0$
of the stochastic perturbation driving the Langevin equation
(\ref{eq2_3}) decays in time as
\begin{equation}
E \left [ (-1)^{\chi(t+\tau)} (-1)^{\chi(\tau)} \right ]
 = e^{- 2 \, \lambda \, t}
\label{eq2_6}
\end{equation}
where $t,\tau \geq 0$.
As $(-1)^{\chi(t)}$ flips between $\pm1$, depending on the
parity of $\chi(t)$, this model of noise
has been referred to as {\em two-state dichotomous
noise}, and since its correlation function  decays
exponentially with time $t$, it has been considered
as one of the  basic prototypes for colored noise.

Albeit these two properties  are certainly important, the
most relevant feature of the Poisson-Kac model is the
finite-propagation velocity of the stochastic perturbation
(provided that $v(x)$ is sufficiently smooth, e.g. $C^1$),
and this reflects into the fact that the trajectories of the
stochastic process $X(t)$ (the small case symbol $x(t)$
indicates a realization of it) are almost everywhere differentiable
functions of time $t$ \cite{giona_bc}.

If $(-1)^{\chi(t)}=1$, eq. (\ref{eq2_3}) becomes
$d x(t)/d t= v(x(t)) + b = v_+(x(t))$, while
if $(-1)^{\chi(t)}=-1$, it reduces to
$d x(t)/d t= v(x(t)) + b = v_-(x(t))$. Therefore,
the statistical description of $X(t)$ can be grounded on
the system of two partial probability density
functions $\{ p^+(x,t), \;
p^-(x,t) \}$, defined as
\begin{equation}
p^\pm(x,t) \, d x = \mbox{Prob} \left [
X(t) \in (x,x+dx) \, , (-1)^{\chi(t)} = \pm 1 \right ]
\label{eq2_7}
\end{equation}
and fulfilling  the system of two first-order
dissipative hyperbolic equations
\begin{eqnarray}
\partial_t p^+(x,t) & = & -\partial_x \left [ v_+(x) \, p^+(x,t) \right ]
-\lambda \, p^+(x,t) + \lambda \, p^-(x,t)
\nonumber \\
\partial_t p^-(x,t) & = & -\partial_x \left [ v_-(x) \, p^-(x,t) \right ]
+\lambda \, p^+(x,t) - \lambda \, p^-(x,t)
\label{eq2_8}
\end{eqnarray}
The partial probability  density functions are also referred
to as the two {\em partial probability waves} of the
Poisson-Kac process.
The finite propagation is ensured by the bounded values
of $v_\pm(x)$, while dissipation is accounted for
by the recombination dynamics $\mp \lambda (p^+- p^-)$
amongst the partial waves.
Indicating with $p(x,t)$ the overall probability density function
for $X(t)$ at time $t$, and with $J_d(x,t)$
the probability flux associated with the
Poissonian stochastic perturbation (i.e., the ``diffusive flux''
in the Poisson-Kac model),
\begin{equation}
p(x,t)= p^+(x,t) + p^-(x,t) \;, \;\;\;
J_d(x,t) = b \, \left [ p^+(x,t) - p^-(x,t) \right ]
\label{eq2_9}
\end{equation}
it follows from eq. (\ref{eq2_8}) that
\begin{eqnarray}
\partial_t p(x,t)  =  - \partial_x \left [ v(x) \, p(x,t) \right ]
-\partial_x J_d(x,t) \nonumber \\
\frac{1}{2 \, \lambda} \, \partial_t J_d(x,t) +
J_d(x,t)  = 
 - \frac{1}{2 \, \lambda} \, \partial_x  \left [ v(x) \, J_d(x,t)
\right ]
- \frac{b^2}{2 \, \lambda} \, \partial_x p(x,t)
\label{eq2_10}
\end{eqnarray}

If $v(x)=0$, the second equation (\ref{eq2_10})
corresponds to a Cattaneo constitutive equation.
This is the reason why the Cattaneo equation, and
the Poisson-Kac process in the absence of a deterministic
bias, i.e., $v(x)=0$, have been considered essentially
as two descriptions (Eulerian vs Lagrangian) of the
same physics. This is not always true as boundary conditions
enters in the picture, and the probabilistic interpretation
of the Cattaneo equation, even in one-dimensional spatial
problems, poses strict and quantitative constraints on the
nature of the admissible boundary conditions \cite{brasiello}.

Nevertheless, even in one spatial dimension, the analogy breaks down
if $v(x) \neq 0$, and does not occur for spatial dimensions $n$ higher than
$1$. More precisely, the Cattaneo constitutive equation
for $n\geq 2$,
\begin{equation}
\tau_c \, \partial_t {\bf J}_d({\bf x},t) +
{\bf J}_d({\bf x},t) = - K \, \nabla p
\label{eq2_11}
\end{equation}
where $\tau_c, K>0$ are constants, ${\bf x}=(x_1,\dots,x_n)$,
${\bf J}_d=(J_{d,1},\dots J_{d,n})$ does not correspond to any
stochastic process. Correspondingly, the substitution of
this constitutive equation into the balance equation
$\partial_t p({\bf x},t)= -\nabla \cdot {\bf J}_d({\bf x},t)$,
leads to an evolution equation for $p({\bf x},t)$, that
starting from a non-negative initial probability density
$p({\bf x},0)=p_0({\bf x}) \geq 0$, may lead at some
time $t>0$
to negative values of $p({\bf x},t)$ \cite{korner}.

The other important property of the Poisson-Kac model is
its asymptotic limit, whenever $b$ and $\lambda$ are let to
diverge $b,\lambda \rightarrow \infty$, keeping fixed
the ratio
\begin{equation}
D_{\rm eff} = \frac{b^2}{2 \, \lambda}
\label{eq2_12}
\end{equation}
Under these conditions,  the second equation (\ref{eq2_10})
reduces to a Fickian form (we use the diction ``Fickian''
to indicate any constitutive equation where the
diffusive flux is proportional to the gradient of the density),
\begin{equation}
J_d(x,t) = - D_{\rm eff} \, \partial_x p(x,t)
\label{eq2_13}
\end{equation}
and the probability balance equation (first equation (\ref{eq2_10}))
converges to the classical parabolic advection-diffusion equation
\begin{equation}
\partial_t p(x,t) = - \partial_x \left [ v(x) \, p(x,t) \right ]
+ D_{\rm eff} \, \partial_x^2 p(x,t)
\label{eq2_14}
\end{equation}
We refer to this limit as the {\em Kac limit}, which implies that:
\begin{itemize}
\item a classical Langevin equation driven by a Wiener process can be
always regarded as the Kac limit of some Poisson-Kac dynamics;
\item  the Poisson-Kac process represents a physically meaningful
(wave-like) mollification of a Wiener process. We return
to this issue in parts II and III, where  the connections
between  the Wong-Zakai theorem \cite{wong_zakai1,wong_zakai2}
and Generalized
Poisson-Kac processes are analyzed.
\end{itemize}


\begin{thebibliography}{160}

\bibitem{gen1} Kleinert H 1988 {\it Gauge fields in condensed matter Vol. 1}
(Singapore: World Scientific)

\bibitem{gen2} Zinn-Justin J 1988 {\it Quantum field theory
and critical phenomena} (Oxford: Oxford Science Publ.)

\bibitem{gen3} Nelson E 1985 {\it Quantum fluctuations} (Princeton: Princeton
University Press)

\bibitem{path1} Feynman R and Hibbs A R 2010 {\it Quantum mechanics
and path integrals} (New York: Dover Publ.)

\bibitem{path2} Kleinert H 2009 {\it Path integrals in quantum mechanics,
statistics, polymer physics and financial markets} (Singapore:
World Scientific)


\bibitem{chandra} Chandrasekhar S 1943 {\it Rev. Mod. Phys.} {\bf 15} 1

\bibitem{ottinger} \"{O}ttinger H C 1996 {\it Stochastic
Processes in Polymeric Fluids} (Berlin: Springer Verlag)

\bibitem{soft} Doi M and  Edwards S F 1988 {\it The theory of polymer
dynamics} (Oxford: Oxford Univ. Press)

\bibitem{de_groot} de Groot S R and Mazur P 1984 {\it Non-equilibrium
thermodynamics} (New York: Dover Publ.)


\bibitem{prigogine} Kondepuni D and Prigogine I 2014 {\it Modern 
Thermodynamics} (Chichister, J. Wiley \& Sons)

\bibitem{falconer} Falconer K J 1986 {\it The geometry of fractal
sets} (Cambridge: Cambridge Univ. Press)

\bibitem{zwanzig} Zwanzig R 1961 {\it Phys. Rev} {\bf 124} 983

\bibitem{cattaneo1} Cattaneo C 1948 {\it Atti Semin. Mat. Fis. Univ. Modena}
{\bf 3} 3

\bibitem{cattaneo2} Cattaneo C 1958 {\it Compt. Rend.} {\bf 247} 431

\bibitem{vernotte} Vernotte P 1958 {\it Comp. Rend.} 246 3154

\bibitem{heat1} Simons S 1973 {\it J. Phys. A} {\bf 6} 1543

\bibitem{heat2} Lebon G, Ruggieri M and Valenti A 2008 {\it J. Phys.
Condens. Matter} {\bf 20} 025223

\bibitem{heat30}  Christov C I and Jordan P M 2005
{\it Phys. Rev. Lett.} {\bf 94} 154301

\bibitem{heat3} Christov C I 2009 {\it Mech. Res. Comm.} {\bf 36} 481

\bibitem{heat4} Cimmelli V A, Sellitto A and Jou D 2010 {\it Phys. Rev. B}
{\bf 81} 054301

\bibitem{heat5} Jou D and Cimmelli V A 2016 {\it Commun. appl. Ind.
Math.} {\bf 7} 196

\bibitem{catappl1} Joseph D D and Preziosi L 1989 {\it Rev. Mod. Phys.}
{\bf 61} 41

\bibitem{catappl2} Ruggeri T, Muracchini A and Seccia L 1990
{\it Phys. Rev. Lett.} {\bf 64} 2640

\bibitem{catappl3} Tung M M, Trujillo M, Molina J L, Rivera M J and Berjano E J
2009 {\it Math. Comp. Modell.} {\bf 50} 665

\bibitem{catappl4} Jou D, Sellitto A and Cimmelli V A 2014
{\it Int. J. Heat Mass Transf.} {\bf 71} 459



\bibitem{statder1}  Olivares-Robles M A and Garcia-Colin L S 1994
{\it Phys. Rev. E} {\bf 50} 2451

\bibitem{statder2} Olivares-Robles M A and Garcia-Colin L S 1996
{\it J. Non-Equilib. Thermodyn.} {\bf 21} 361

\bibitem{pawula}  Pawula R F 1967 {\it Phys. Rev.}  {\bf 162} 186

\bibitem{statder3} Zhang Z M, Bright T J and Peterson G P 2011
{\it Nano. Micro. Thermophys. Eng.} {\bf 15} 220.

\bibitem{statder4} Metzler R and Compte A 1999 {\it Physica A} {\bf 268}
454

\bibitem{grad}   Grad H 1949 {\it Comm. Pure Appl. Math.} {\bf 2} 331

\bibitem{muller1} M\"uller I 1967 {\it Z. Phys.} {\bf 198} 329


\bibitem{muller2} M\"uller I and Ruggeri T 2013 {\it Rational
extended thermodynamics} (New York: Springer Science \& Business Media)

\bibitem{muller3} M\"uller I 1999 {\it Living Rev. Relativity} {\bf 2} 1


\bibitem{jou0} Garcia-Colin L S, Lopez de Haro M, Rodriguez R F, 
Casas-Vazquez J and Jou D 1984 {\it J. Stat. Phys.} {\bf 37} 465

\bibitem{jou1}   Jou D, Casas-Vazquez J and Lebon G 1988 {\it Rep. Prog.
Phys.} {\bf 51} 1105


\bibitem{jou2} Jou D, Casas-Vazquez J and Lebon G 1999 {\it Rep. Prog.
Phys.} {\bf 62} 1035


\bibitem{jou3} Jou D, Casas-Vazquez J and Lebon G 1996
{\it Extended irreversible thermodynamics} (Berlin: Springer
Verlag)

\bibitem{kac}  Kac M 1974 {\it Rocky Mount. J. Math.} {\bf 4} 497

\bibitem{goldstein} Goldstein S 1951 {\it Quart. J. Mech. Appl. Math.} {\bf 4}
129


\bibitem{dicho1} Sancho J  1984 {\it J. Math. Phys.} {\bf 25} 354
\bibitem{dicho2} Pawula R F 1987 {\it Phys. Rev. A} {\bf 35} 3102

\bibitem{dicho3} Masoliver J, Porr\'{a} J M and Weiss G H 1993  {\it Phys. Rev. E}
{\bf 48} 939
\bibitem{dicho4} Masoliver J and Weiss G H 1994  {\it Phys. Rev. E} {\bf 49}
3852
\bibitem{dicho5} Dean Astumian R and Bier M  1994 {\it Phys. Rev. Lett.}
{\bf 72} 1766
\bibitem{dicho6} Porr\'{a} J M and Lindenberg K 1995 {\it Phys. Rev. E} 
{\bf 52} 409
\bibitem{dicho7} Fulinski A 1995 {\it Phys. Rev. E} {\bf 52} 4523
\bibitem{dicho8} Masoliver J and Weiss G H 1996 {\it Eur. J. Phys.} 190
\bibitem{dicho9} Reimann P and Elston T C 1996 {\it Phys. Rev. Lett.}
{\bf 77} 5328
\bibitem{dicho10} Boguna M, Porr\'{a} J M and Masoliver  J 1999
{\it Phys. Rev. E} {\bf 59} 6517
\bibitem{dicho11} Laio F, Ridolfi L and D'Odorico P 2008 {\it Phys. Rev.
E} {\bf 78} 031137
\bibitem{color1} Arnold L, Horsthemke W and Levefer R 1978
{\it Z. Phys. B} {\bf 29} 367
\bibitem{color2} Kus M, Wajnryb E and Wodkiewicz K  1991 {\it Phys.
Rev. A} {\bf 43} 4167
\bibitem{color3} H\"anggi P and Jung P 1995 {\it Adv. Chem. Phys.} {\bf 89}
239
\bibitem{color4} Kim S, Park S H and Ryu C S 1998 {\it Phys. Rev. E}
{\bf 58} 7994
\bibitem{weiss} Weiss G H 2002 {\it Physica A} {\bf 311} 381
\bibitem{bena} Bena I 2006 {\it Int. J. Mod. Phys. B} {\bf 20} 2825
\bibitem{horsthemke} Horsthemke W and Lefever  R 2006 {\it
Noise-Induced Transitions} (Berlin: Springer Verlag) 
\bibitem{nitrans1} Kitahara K, Horsthemke W and Lefever R 1979
{\it Phys. Lett. A} {\bf 70} 377
\bibitem{nitrans2} Kitahara K, Horsthemke W, Lefever R and Inaba Y
1980 {\it Prog. Theor. Phys.} {\bf 64} 1233
\bibitem{nitrans3} Horsthemke W, Doering C R,  Lefever R, and Chi A S,
1985 {\it Phys. Rev. A} {\bf 31} 1123
\bibitem{nitrans4} M\"uller R, and Behn U 1990 {\it Z. Phys. B} {\bf 78}
229
\bibitem{gaveau} Gaveau B, Jacobson T, Kac M and Schulman L S 1984
{\it Phys. Rev. Lett.} {\bf 53} 419
\bibitem{dirac1} Kudo T and Ohba I 2002 {\it Praman} {\bf 59} 413

\bibitem{dirac2} Balakrishnan V and Lakshmibala S 2005 {\it New J. Phys.}
{\bf 7} 11
\bibitem{dirac3} Srinivasan S K and Sudarshan E C G 1996 {\it J. Phys. A}
{\bf 29} 5181
\bibitem{jona1} Beccaria M, Presilla C, De Angelis G F and Jona-Lasinio G
1999 {\it Europhys. Lett.} {\bf 48} 243
\bibitem{jona2} Beccaria M, Presilla C, De Angelis G F and Jona-Lasinio G
2001 {\it Int. J. Mod. Phys. B} {\bf 15} 1740
\bibitem{korner} K\"orner C and Bergmann H W 1998 {\it Appl. Phys. A}
{\bf 67} 397
\bibitem{criticism0} Bright T J and Zhang Z M 2009 {\it J. Thermophys.
Heat Transf.} {\bf 23} 601
\bibitem{criticism1} Vavruch I 2002  {\it Chemick\'e listy} {\bf 96}
271
\bibitem{criticism_extended} Dunkel J and H\"anggi P 2009 {\it Phys.
Rep.} {\bf 471} 1
\bibitem{zhuk} Zhukovsky K 2016 {\it Int. J. Heat Mass Transf.} {\bf 98}
523
\bibitem{kolesnik1} Kolesnik A D 2008 {\it J. Stat. Phys.} {\bf 131} 1039
\bibitem{kolesnik2} Kolesnik A D and Pinsky M A 2011
{\it J. Stat. Phys.} {\bf 142} 828
\bibitem{masoliver1} Masoliver J, Porr\`a J M and Weiss G H 1992
{\it Physica A} {\bf 192} 593
\bibitem{masoliver2} Masoliver J, Porr\`a J M and Weiss G H 1993
{\it Physica A} {\bf 193} 469
\bibitem{boguna} Boguna M, Porr\`{a} J P and Masoliver J 1998 
{\it Phys. Rev. E} {\bf 58} 6992
\bibitem{colin} Godoy S and Garcia-Colin  L S 1997 {\it Phys. Rev. E} {\bf 55}
2127
\bibitem{part2}   Giona M, Brasiello A and Crescitelli S 2016
Stochastic foundations of
undulatory transport phenomena:  Generalized Poisson-Kac processes - Part II
 Irreversibility, Norms and Entropies, submitted to {\it J. Phys. A}
\bibitem{part3} Giona M, Brasiello A and Crescitelli S 2016   Stochastic foundations of
undulatory transport phenomena: Generalized Poisson-Kac processes  - Part III
 Extensions and applications to kinetic theory and transport, submitted to
{\it J. Phys. A}


\bibitem{rosenau} Rosenau P 1993 {\it Phys. Rev. E} {\bf 48} R655

\bibitem{giona_gpk} Giona M, Brasiello A and Crescitelli S 2016
{\it J. Non-Equil. Thermodyn.} {\bf 41} 107

\bibitem{kacprog1} Kac M 1956 Foundations of kinetic theory,
in {\em Proc. 3rd Berkeley Symp. Math. Stat. Prob.},
J. Neyman (Ed.), Univ of California, vol. 3, 171

\bibitem{kacprog2} Mischler S and Mouhot C 2013 {\it Invent. Math.}
{\bf 193} 1
\bibitem{kacprog3} Michler S 2013  Kac's chaos and Kac's program,
{\it arXiv preprint} arXiv:1311.7544

\bibitem{lattice_boltzmann}  Doolen  G et al. (Eds.) 1990
{\it Lattice Gas Methods for Partial Differential Equations}
(Menlo Park: Addison-Wesley) 
\bibitem{dbe1} Cabennes H, Gatisgnol R and Luo L S 2003
{\it The discrete Boltzmann equation} (Berkeley: University of California)
\bibitem{dbe2} Palczewski A and Schneider J 1998
{\it J. Stat. Phys.}  {\bf 1998} 307
\bibitem{dbe3} Vinerean M C, Windf\"all A and Bobylev A V 2010
{\it Nuovo Cimento} {\bf 33} 257
\bibitem{brasiello} Brasiello A, Crescitelli S and Giona M 2016
{\it Physica A} {\bf 449} 176
\bibitem{giona_spinorial} Giona M 2016 Covariance and
spinorial  statistical description of simple
relativistic stochastic kinematics, in preparation 
\bibitem{kolesnik3} Kolesnik A  D and Turbin  A F 1998
{\it Stoch. Proc. Appl.} {\bf 75} 67
\bibitem{guyer1} Guyer R A and Krumhansl J A 1966 {\it Phys. Rev.} {\bf 148}
766
\bibitem{guyer2} Guyer R A and Krumhansl J A 1966 {\it Phys. Rev.} {\bf 148}
778
\bibitem{guyer3} Cimmelli V A and Frischmuth K 2007  {\it Physica B}
{\bf 400} 257
\bibitem{plykhin} Plyukhin A V 2010 {\it Phys. Rev. E} {\bf 81} 021113
\bibitem{seneta} Seneta E 2006 {\it Non-negative matrices and Markov Chains}
(New York: Springer Science \& Business Media)

\bibitem{giona_adjoint} Giona M, Brasiello A and Crescitelli S 2016
Markovian nature, completeness, regularity and correlation properties
of Generalized Poisson-Kac processes, submitted to {\it J. Sta. Mech.}

\bibitem{brenner} Brenner H and Edwards D A 1993
{\it Macrotransport Processes} (Boston: Butterworth-Heinemann)

\bibitem{giona_bc} Giona M, Brasiello A and Crescitelli S 2016 {\it Physica
A} {\bf 450} 148
\bibitem{giona_epl} Giona M, Brasiello A and Crescitelli S 2015
{\it Europhys. Lett.} 112 30001
\bibitem{wong_zakai1}   Wong E and Zakai M 1965 {\it Int. J. Eng. Sci.}
{\bf 3} 213
\bibitem{wong_zakai2}  Wong E and Zakai M 1965 {\it Ann. Math. Stat.} {\bf 36}
1560


\end{thebibliography}
\end{document}